\documentclass[trackchanges,twocolumn]{aastex701}

\usepackage{amsmath}
\usepackage{changepage}
\usepackage{graphicx}
\usepackage{subcaption}
\usepackage{lineno}
\usepackage{float}
\usepackage{xcolor}
\begin{document}

\title{SN\,2024aecx: a fast-evolving Type IIb supernova with a prominent shock-cooling peak}

\shorttitle{SN\,2024aecx}
\shortauthors{Xi et al.}

\correspondingauthor{Ning-Chen Sun}

\author{Qiang Xi}
\email{xiqiang23@mails.ucas.ac.cn}
\affiliation{School of Astronomy and Space Science, University of Chinese Academy of Sciences, Beijing 100049, China}
\affiliation{National Astronomical Observatories, Chinese Academy of Sciences, Beijing 100101, China}

\author{Ning-Chen Sun}
\email[show]{sunnc@ucas.ac.cn}
\affiliation{School of Astronomy and Space Science, University of Chinese Academy of Sciences, Beijing 100049, China}
\affiliation{National Astronomical Observatories, Chinese Academy of Sciences, Beijing 100101, China}
\affiliation{Institute for Frontiers in Astronomy and Astrophysics, Beijing Normal University, Beijing, 102206, China}

\author[0000-0001-5200-3973]{David Aguado}
\email{david.aguado@iac.es}
\affiliation{Instituto de Astrof\'{\i}sica de Canarias, V\'{\i}a   L\'actea, 38205 La Laguna, Tenerife, Spain}
\affiliation{Universidad de La Laguna, Departamento de Astrof\'{\i}sica,  38206 La Laguna, Tenerife, Spain}

\author{Ismael P\'erez-Fournon}
\email{ipf@iac.es}
\affiliation{Instituto de Astrof\'{\i}sica de Canarias, V\'{\i}a   L\'actea, 38205 La Laguna, Tenerife, Spain}
\affiliation{Universidad de La Laguna, Departamento de Astrof\'{\i}sica,  38206 La Laguna, Tenerife, Spain}

\author[0000-0002-5391-5568]{Fr\'ed\'erick Poidevin}
\email{frederick.poidevin@iac.es}
\affiliation{Instituto de Astrof\'{\i}sica de Canarias, V\'{\i}a   L\'actea, 38205 La Laguna, Tenerife, Spain}
\affiliation{Universidad de La Laguna, Departamento de Astrof\'{\i}sica,  38206 La Laguna, Tenerife, Spain}

\author[0000-0002-8402-3722]{Junjie Jin}
\email{jjjin@bao.ac.cn}
\affiliation{National Astronomical Observatories, Chinese Academy of Sciences, Beijing 100101, China}

\author{Yiming Mao}
\email{maoym@bao.ac.cn}
\affiliation{National Astronomical Observatories, Chinese Academy of Sciences, Beijing 100101, China}
\affiliation{School of Astronomy and Space Science, University of Chinese Academy of Sciences, Beijing 100049, China}

\author[0000-0002-3651-0681]{Zexi Niu}
\email{nzx@nao.cas.cn}
\affiliation{School of Astronomy and Space Science, University of Chinese Academy of Sciences, Beijing 100049, China}
\affiliation{National Astronomical Observatories, Chinese Academy of Sciences, Beijing 100101, China}

\author{Beichuan Wang}
\email{wangbc@bao.ac.cn}
\affiliation{National Astronomical Observatories, Chinese Academy of Sciences, Beijing 100101, China}
\affiliation{School of Astronomy and Space Science, University of Chinese Academy of Sciences, Beijing 100049, China}

\author{Yu Zhang}
\email{yzhang@bao.ac.cn}
\affiliation{National Astronomical Observatories, Chinese Academy of Sciences, Beijing 100101, China}

\author[0000-0003-1637-267X]{Kuntal Misra}
\email{kuntal@aries.res.in}
\affiliation{Aryabhatta Research Institute of Observational Sciences, Nainital-263001, India}

\author{Divyanshu Janghel}
\email{divyanshu@aries.res.in}
\affiliation{Aryabhatta Research Institute of Observational Sciences, Nainital-263001, India}
\affiliation{Mahatma Jyotiba Phule Rohilkhand University, Bareilly-243006, India}

\author{Justyn R. Maund}
\email{justyn.maund@rhul.ac.uk}
\affiliation{Department of Physics, Royal Holloway, University of London, Egham, TW20 0EX, United Kingdom}

\author[0000-0002-4870-9436]{Amit Kumar}
\email{Amit.Kumar@rhul.ac.uk}
\affiliation{Department of Physics, Royal Holloway, University of London, Egham, TW20 0EX, United Kingdom}

\author{Samaporn Tinyanont}
\email{samaporn@narit.or.th}
\affiliation{National Astronomical Research Institute of Thailand, 260 Moo 4, Donkaew, Maerim, Chiang Mai, 50180, Thailand}

\author[0000-0002-8708-0597]{Liang-Duan Liu}
\email{liuld@ccnu.edu.cn}
\affiliation{Institute of Astrophysics, Central China Normal University, Wuhan 430079, China}
\affiliation{Education Research and Application Center, National Astronomical Data Center, Wuhan 430079, China}
\affiliation{Key Laboratory of Quark and Lepton Physics (Central China Normal University), Ministry of Education, Wuhan 430079, China}

\author{Yu-Hao Zhang}
\email{zhang-yh@mails.ccnu.edu.cn}
\affiliation{Institute of Astrophysics, Central China Normal University, Wuhan 430079, China}
\affiliation{Education Research and Application Center, National Astronomical Data Center, Wuhan 430079, China}
\affiliation{Key Laboratory of Quark and Lepton Physics (Central China Normal University), Ministry of Education, Wuhan 430079, China}

\author{Bhavya Ailawadhi}
\email{bhavya@prl.res.in}
\affiliation{Aryabhatta Research Institute of Observational Sciences, Nainital-263001, India}
\affiliation{Astronomy \& Astrophysics Division, Physical Research Laboratory, Ahmedabad 380009, Gujarat, India}

\author{Monalisa Dubey}
\email{monalisa@aries.res.in}
\affiliation{Aryabhatta Research Institute of Observational Sciences, Nainital-263001, India}
\affiliation{Mahatma Jyotiba Phule Rohilkhand University, Bareilly-243006, India}

\author[0000-0003-0292-4832]{Zhen Guo}
\email{zhen.guo@uv.cl}
\affiliation{Instituto de F{\'i}sica y Astronom{\'i}a, Universidad de Valpara{\'i}so, ave. Gran Breta{\~n}a, 1111, Casilla 5030, Valpara{\'i}so, Chile}
\affiliation{Millennium Institute of Astrophysics, Nuncio Monse{\~n}or Sotero Sanz 100, Of. 104, Providencia, Santiago, Chile}

\author{Anshika Gupta}
\email{anshika@aries.res.in}
\affiliation{Aryabhatta Research Institute of Observational Sciences, Nainital-263001, India}
\affiliation{Department of Physics, Indian Institute of Technology Roorkee, Roorkee 247667, India}

\author[0000-0001-6139-7660]{Min He}
\email{hemin@bao.ac.cn}
\affiliation{National Astronomical Observatories, Chinese Academy of Sciences, Beijing 100101, China}

\author{Dhruv Jain}
\email{dhruvjain@aries.res.in}
\affiliation{Aryabhatta Research Institute of Observational Sciences, Nainital-263001, India}
\affiliation{Mahatma Jyotiba Phule Rohilkhand University, Bareilly-243006, India}

\author{Debalina Kar}
\email{debalina@aries.res.in}
\affiliation{Aryabhatta Research Institute of Observational Sciences, Nainital-263001, India}
\affiliation{Department of Physics, Indian Institute of Technology Roorkee, Roorkee 247667, India}

\author{Wenxiong Li}
\email{liwx@bao.ac.cn}
\affiliation{National Astronomical Observatories, Chinese Academy of Sciences, Beijing 100101, China}

\author[0000-0002-3464-0642]{Joe D. Lyman}
\email{J.D.Lyman@warwick.ac.uk}
\affiliation{Department of Physics, University of Warwick, Coventry, CV4 7AL, UK}

\author[]{Haiyang Mu}
\email{hymu@nao.cas.cn}
\affiliation{National Astronomical Observatories, Chinese Academy of Sciences, Beijing 100101, China}
\affiliation{School of Astronomy and Space Science, University of Chinese Academy of Sciences, Beijing 100049, China}

\author{Kumar Pranshu}
\email{pranshu@aries.res.in}
\affiliation{Aryabhatta Research Institute of Observational Sciences, Nainital-263001, India}
\affiliation{Department of Applied Optics and Photonics, University of Calcutta, Kolkata, 700106, India}

\author[0009-0002-0735-3274]{Xinxiang Sun}
\email{sunxx@nao.cas.cn}
\affiliation{National Astronomical Observatories, Chinese Academy of Sciences, Beijing 100101, China}
\affiliation{School of Astronomy and Space Science, University of Chinese Academy of Sciences, Beijing 100049, China}

\author[0000-0002-1094-3817]{Lingzhi Wang}
\email{wanglingzhibnu@gmail.com}
\affiliation{Chinese Academy of Sciences South America Center for Astronomy (CASSACA), National Astronomical Observatories, CAS, Beijing 100101, China}
\affiliation{Departamento de Astronom{\'i}a, Universidad de Chile, Las Condes, 7591245 Santiago, Chile}

\author{Sarvesh Kumar Yadav}
\email{sarvesh@aries.res.in}
\affiliation{Aryabhatta Research Institute of Observational Sciences, Nainital-263001, India}
\affiliation{Department of Applied Optics and Photonics, University of Calcutta, Kolkata, 700106, India}

\author{Yi-Han Zhao}
\email{zhaoyihan20@mails.ucas.ac.cn}
\affiliation{School of Astronomy and Space Science, University of Chinese Academy of Sciences, Beijing 100049, China}
\affiliation{National Astronomical Observatories, Chinese Academy of Sciences, Beijing 100101, China}

\author[0000-0001-6637-6973]{Jie Zheng}
\email{jiezheng@nao.cas.cn}
\affiliation{National Astronomical Observatories, Chinese Academy of Sciences, Beijing 100101, China}

\author{Yinan Zhu}
\email{ynzhu@nao.cas.cn}
\affiliation{National Astronomical Observatories, Chinese Academy of Sciences, Beijing 100101, China}

\author{David L\'opez Fern\'andez-Nespral}
\email{david.lopez@iac.es}
\affiliation{Instituto de Astrof\'{\i}sica de Canarias, V\'{\i}a   L\'actea, 38205 La Laguna, Tenerife, Spain}
\affiliation{Universidad de La Laguna, Departamento de Astrof\'{\i}sica,  38206 La Laguna, Tenerife, Spain}

\author[0000-0003-4603-1884]{Alicia L\'opez Oramas}
\email{alicia.lopez@iac.es}
\affiliation{Instituto de Astrof\'{\i}sica de Canarias, V\'{\i}a   L\'actea, 38205 La Laguna, Tenerife, Spain}
\affiliation{Universidad de La Laguna, Departamento de Astrof\'{\i}sica,  38206 La Laguna, Tenerife, Spain}

\author{Yanan Wang}
\email{wangyn@bao.ac.cn}
\affiliation{National Astronomical Observatories, Chinese Academy of Sciences, Beijing 100101, China}

\author{Klaas Wiersema}
\email{k.wiersema@herts.ac.uk}
\affiliation{Centre for Astrophysics Research, University of Hertfordshire, Hatfield, AL10 9AB, UK}

\author{Jifeng Liu}
\email{jfliu@nao.cas.cn}
\affiliation{National Astronomical Observatories, Chinese Academy of Sciences, Beijing 100101, China}
\affiliation{School of Astronomy and Space Science, University of Chinese Academy of Sciences, Beijing 100049, China}
\affiliation{Institute for Frontiers in Astronomy and Astrophysics, Beijing Normal University, Beijing, 102206, China}

\begin{abstract}

SN\,2024aecx is a nearby ($\sim$11\,Mpc) Type~IIb SN discovered within $\sim$1\,d after explosion. In this paper we report high-cadence photometric (typically 0.5$\sim$1 day) and spectroscopic follow-up observations, conducted from as early as 0.27\,d post discovery out to the nebular phase at 158.4\,d. We analyze the environment of SN\,2024aecx and derive a new distance (11.3$\pm$1.1\,Mpc), metallicity and host extinction. The light curve exhibits a hot and luminous shock-cooling peak at the first few days, followed by a main peak with very rapid post-maximum decline. The earliest spectra are blue and featureless, while from 2.3\,d after discovery prominent P-Cygni profiles emerge. At nebular phase, the emission lines exhibit asymmetric and double-peaked profiles, indicating asphericity and/or early dust formation in the ejecta. Nebular spectral modelling indicates a blueshifted O-rich clump moving toward observer, and the \([\text{O\,\textsc{i}}]/[\text{Ca\,\textsc{ii}}]\) line ratio suggests an intermediate-mass progenitor. We simulated the progenitor and explosion using a two-component model of shock cooling and radioactive $^{56}$Ni heating; our model favors an extended, low-mass H-rich envelope with \( M_{\mathrm{e}} = 0.04\pm{0.01}\, M_{\odot} \) and a low ejecta mass of \( M_{\mathrm{ej}} = 1.55^{+0.18}_{-0.14} \, M_{\odot} \). And the nebular-phase spectra and light-curve modelling both suggest that it most likely originated from an intermediate-mass binary progenitor system. The comprehensive monitoring of SN\,2024aecx, coupled with the detailed characterization of its local environment, establishes it as a benchmark event for probing the progenitors and explosion mechanisms of Type~IIb SNe.

\end{abstract}



\section{Introduction}
\label{sec:intro}

Supernovae (SNe) are the catastrophic endpoints of stellar evolution. They inject energy, momentum, and freshly synthesized elements into the interstellar medium, thereby shaping star formation, feedback, and chemical enrichment in galaxies. Spectroscopically, SNe are divided into Type~I, which lack conspicuous hydrogen Balmer features near maximum light, and Type~II, which display prominent hydrogen lines \citep{Minkowski1941,Filippenko1997}. Type~Ia events are the thermonuclear disruptions of carbon–oxygen white dwarfs, while the other spectral types arise from the core collapse of massive stars with initial masses $\gtrsim8$–$10~M_\odot$ \citep{Filippenko1997,Smartt2009}. Within the core-collapse (CC) SN family, Type~II SNe are hydrogen-rich, Type~Ib lack hydrogen but show helium, and Type~Ic show neither hydrogen nor helium. The progressive loss of specific spectral features reflects different degrees of envelope stripping by line-driven winds and/or binary mass transfer of their progenitors \citep{Yoon2010,Langer2012,Smith2014,Sun2023b}. Thus, these types (Ib, Ic, and IIb) of SNe are also referred to as stripped-envelope (SE) SNe.

Type~IIb is a transitional SN class which exhibits hydrogen lines only at early times. Within days to weeks, their hydrogen lines disappear and helium features begin to dominate (i.e. showing a Type~II to Type~Ib transformation) \citep{Filippenko1988,Arcavi2011}. They often exhibit a short-lived, blue shock-cooling peak in their early light curves, which can sensitively constrain the progenitor radius and the mass of any residual hydrogen layer \citep{Schmidt1993J, Ergon20142011dh, Tartaglia2016gkg}. Type~IIb events are therefore powerful probes of partial envelope removal and pre-explosion mass loss.

\begin{figure}
    \centering

    \includegraphics[width=1\linewidth]{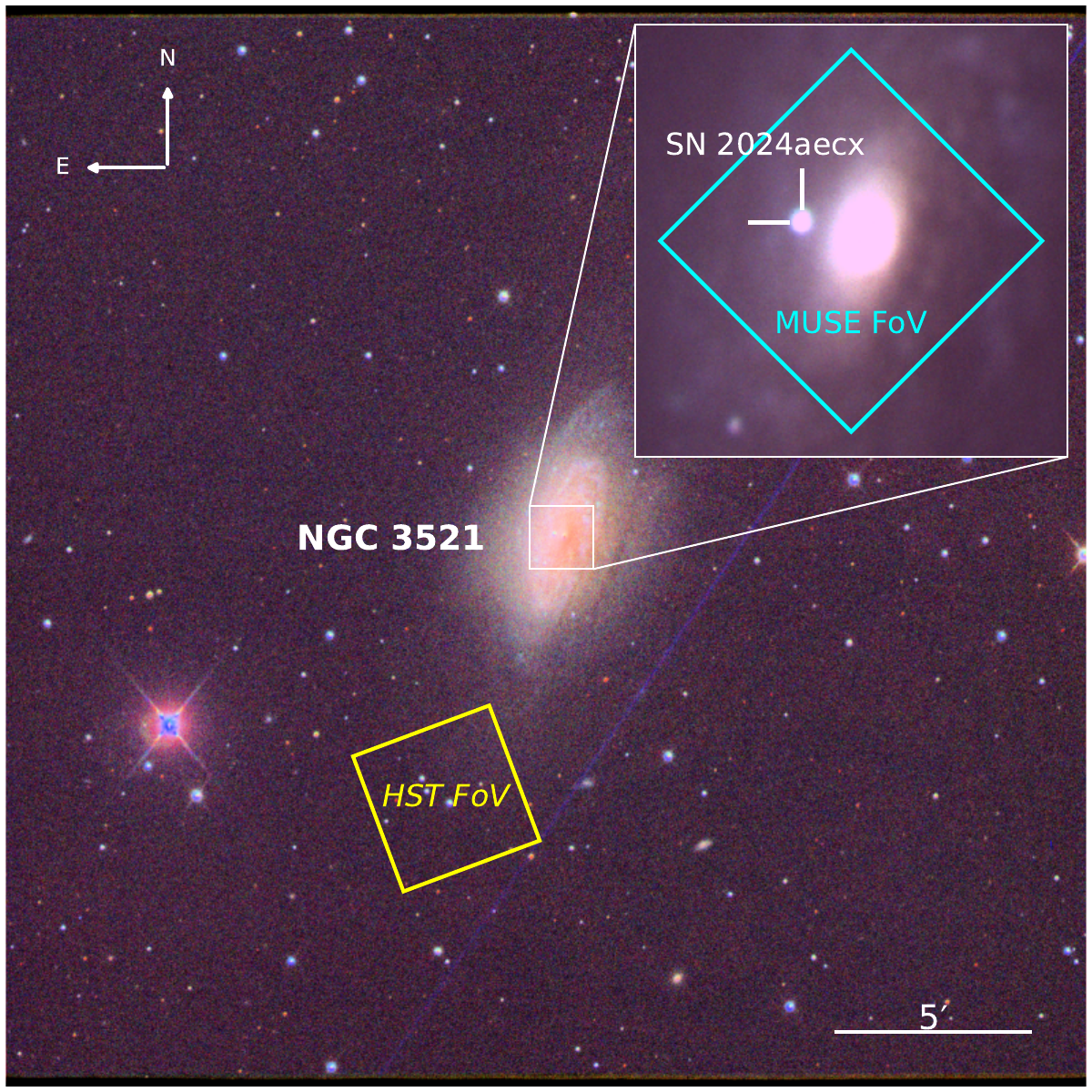}
    \caption{SN 2024aecx and its host galaxy NGC 3521. Shown is an RGB composite assembled from LCO $g$-, $r$-, and $i$-band exposures; the inset at upper right highlights the enlarged region around the galaxy’s nucleus, with SN\,2024aecx marked by a cross. The yellow and cyan squares denote the HST/ACS and VLT/MUSE fields of view, respectively. North is up and East is to the left.}
    \label{fig:host}
\end{figure}

SN\,2024aecx is a nearby SN in the late-type spiral NGC\,3521 (Fig.~\ref{fig:host}) at a redshift of only $z=0.002665$ and a distance of only $D=11.3$\,Mpc (see Section~\ref{sec:dist}). It was discovered by ATLAS on 2024~December~16th, ~13{:}22{:}23.808~UT (MJD$_0=60660.56$) \citep{TNSAN371}. The last non-detections by ATLAS and ZTF are within $\sim$1\,day before discovery, thereby tightly constraining the explosion epoch \citep{TNSAN374}. Early follow-up spectra obtained on 2024~December~17 by the SCAT and DLT40 teams suggested a Type~IIb classification, while a DLT40 spectrum on December~19 suggested a Type~Ic classification \citep{TNSCR4943,TNSCR4947,TNSCR4974}. A comprehensive analysis by \citet{Zou2025}—combining dense multi-band photometry with low-resolution spectroscopy—shows an early, short-lived peak arising from the shock-cooling of a low-mass hydrogen envelope; their \textsc{synapps} spectral synthesis identifies weak H\,\textsc{i} that fades by $\sim$30~d after explosion, confirming the Type~IIb classification.

Therefore, as a nearby Type~IIb SN discovered at an exceptionally early phase and exhibiting a prominent shock-cooling peak, SN\,2024aecx is a golden target for studying the progenitor and explosion physics of Type~IIb SNe. Soon after discovery, we carried out rapid photometric and spectroscopic follow-up observations. Photometry was performed in both ultraviolet (UV) and optical wavelengths and the first photometric data points were acquired at only $t=0.26$\,d after discovery; we densely sampled the light curves to probe the fast evolving early peak as well as the latter evolution. We obtained the first spectrum at only $t=0.3$\,d and continued the observations through the nebular phase out to $t=158.4$\,d. In this paper we present the results and analysis in detail.

Section~\ref{sec:obs} describes our observations of SN\,2024aecx, and Section~\ref{sec:host} analyzes its host galaxy, NGC\,3521. Section~\ref{sec:lightcurve} presents the light curves, and Section~\ref{sec:spectroscopy} shows the spectral sequence. In Section~\ref{sec:model}, we model the progenitor and bolometric light curve of SN\,2024aecx. We summarize our conclusions in Section~\ref{sec:conclusion}. All epochs in this paper are relative to discovery unless otherwise specified.


\section{Observations}
\label{sec:obs}

\subsection{Photometry}

\begin{deluxetable}{ccccc}
    \tablecaption{UV-Optical photometry of SN\,2024aecx. Magnitudes are reported in the default photometric systems for each filter set: Johnson–Cousins $UBVRI$ and Swift/UVOT bands are calibrated in the Vega system, whereas SDSS $ugriz$, ATLAS $c$ and $o$, and the GOTO $L$ band are defined in the AB system. The photometric uncertainties are given in $1\sigma$. Only the first 10 rows are shown here; the full table is available online. 
    \label{tab:photometry}}
    \tablehead{
    \colhead{MJD} & \colhead{Instrument} & \colhead{Filter} & \colhead{Magnitude} & \colhead{Error}
    }
    \startdata
    60660.56 & ATLAS       & o         & 14.71 & 0.01 \\
    60660.82 & UCAS-70     & B         & 14.95 & 0.08 \\
    60660.82 & UCAS-70     & R         & 14.68 & 0.11 \\
    60660.82 & UCAS-70     & V         & 14.68 & 0.10 \\
    60660.82 & UCAS-70     & I         & 14.63 & 0.09 \\
    60660.83 & Swift/UVOT  & Swift B   & 15.25 & 0.06 \\
    60660.83 & Swift/UVOT  & Swift U   & 13.98 & 0.05 \\
    60660.83 & Swift/UVOT  & UVW1      & 14.02 & 0.06 \\
    60660.84 & Swift/UVOT  & Swift V   & 15.14 & 0.08 \\
    60660.84 & Swift/UVOT  & UVW2      & 14.64 & 0.07 \\
    \enddata
\end{deluxetable}

\subsubsection{Ground-Based Telescopes}
We conducted a multi-band photometric monitoring campaign of SN\,2024aecx in the $u,g,r,i,z$ and $U,B,V,R,I$ bands with six facilities: the Xinglong 60\,cm telescope (XL-60; \citealt{Mu2024xl60}), the 70\,cm telescope of the University of Chinese Academy of Sciences (UCAS-70; \citealt{Mao2025ucas70}) the 1.3\,m Devasthal Fast Optical Telescope (DFOT; \citealt{Joshi2022DFOT}), the 2.0\,m Liverpool Telescope (LT; \citealt{Steele2001lt}), the Thai Robotic Telescope network\footnote{\url{https://trt.narit.or.th/}} (TRT), and the Las Cumbres Observatory Global Telescope network (LCO; \citealt{Brown2013LCO}). The first data points were obtained as early as $t=0.26$\,d and we used a very high cadence to densely sample the light curves. When SN\,2024aecx had become substantially faint at late times, we obtained deep-field images in all observed bands with each facility to serve as template frames.

We performed image subtraction using the \textsc{hotpants} software \citep{Becker2015} and carried out point-spread-function (PSF) photometry on the difference images with the \textsc{autophot} package \citep{Brennan2024}. Photometric calibration was achieved using zero-point values derived from the template frames. For the $u,g,r,i,z$ bands, we used stars from the SDSS catalog \citep{Ahn2014} for calibration. For the $B$, $V$, $R$, and $I$ bands, we calibrated against the APASS catalog \citep{Henden2015}, adopting the transformation equations of \citet{Lupton2005}. For the Johnson $U$ band, we calibrated using SDSS catalog transformed via the equations of \citet{Jester2005}. A list of photometry is provided in Table~\ref{tab:photometry}.

\subsubsection{The Swift Satellite}

The UltraViolet/Optical Telescope (UVOT; \citealt{Roming2005}) onboard the Neil Gehrels Swift Observatory \citep{Gehrels2004} conducted multi-epoch follow-up observations of SN\,2024aecx (ObsID\,00018989) in the $UVW2$, $UVM2$, $UVW1$, $U$, $B$ and $V$ filters starting from $t=0.27$\,d. Before explosion there had been archival observations of NGC\,3521 in the $UVW2$, $UVM2$ and $UVW1$ filters  (ObsID\,00084364); we co-added these pre-explosion images as deep templates in the UV filters. After the SN had faded significantly, we obtained Swift/UVOT Target-of-Opportunity exposures in the $U$, $B$ and $V$ bands to serve as optical templates. We performed aperture photometry with \texttt{HEASoft}/\texttt{uvotsource} using a $5''$ radius and subtracted the host-galaxy background measured with the template images. The results of photometry can be found in Table~\ref{tab:photometry}.

\subsubsection{Forced Photometry of Transient Surveys}

Via the public forced-photometry services, we retrieved the ZTF $g$- and $r$-band light curves and the ATLAS light curves in the cyan ($c$; 420–650 nm) and orange ($o$; 560–820 nm) bands. In addition, we also obtained the $L$-band light curve of SN\,2024aecx from GOTO (the Gravitational-wave Optical Transient Observer), which is a wide-field survey facility designed to perform rapid optical follow-up of gravitational-wave events and other transient phenomena \citep{Steeghs2022}. The $L$ band is a broad optical filter employed by GOTO, covering approximately 4000–7000\,\AA. By integrating over a wide spectral range, the $L$ band maximizes sensitivity and is particularly well suited for the early detection and monitoring of optical transients.

\subsection{Spectroscopy}

\begin{deluxetable}{ccccccc}
\tablecaption{Spectroscopic observation log of SN 2024aecx.\label{tab:spec}}
\tablewidth{0pt}
\tablehead{
  \colhead{MJD} & \colhead{Epoch$^{a}$ (d)} & \colhead{Filter} & \colhead{Grism/VPH} & \colhead{Spectral range (\AA)} & \colhead{Resolution (R)$^{e}$} & \colhead{Telescope/Instrument}
}
\tablecolumns{7}
\startdata
60660.9  & 0.3   & 385LP & G4                      & 3700--8800                     & \(\sim1300\)         & XL-216/BFOSC \\
60661.2  & 0.7   &       & R2500U                  & 3440--4610                     & \(\sim2500\)        & GTC/OSIRIS+ \\
         &       &       & R2500V                  & 4500--6000                     & \(\sim2500\)        &  \\
         &       &       & R1000R                  & 5100--10000                    & \(\sim1000\)        &  \\
60661.6  & 1.0   &       & B480+\_G5309            & 3700--7500                     & \(\sim1500\)         & Gemini-N/GMOS$^b$ \\
60661.6  & 1.0   &       & B+R                     & 3200--5200, 5100--9700        & \(\sim1000,~\sim1300\) & UH88/SNIFS$^c$ \\
60661.8  & 1.2   & 385LP & G4                      & 3700--8800                     & \(\sim1300\)         & XL-216/BFOSC \\
60662.9  & 2.3   & 385LP & G4                      & 3700--8800                     & \(\sim1300\)         & XL-216/BFOSC \\
60663.6  & 3.0   &       & B480+\_G5309            & 3750--7500                     & \(\sim1500\)         & Gemini-N/GMOS$^d$ \\
60666.9  & 6.3   & 385LP & G4                      & 3700--8800                     & \(\sim1300\)         & XL-216/BFOSC \\
60667.9  & 7.3   & 385LP & G4                      & 3700--8800                     & \(\sim1300\)         & XL-216/BFOSC \\
60670.9  & 10.3  & 385LP & G4                      & 3700--8800                     & \(\sim1300\)         & XL-216/BFOSC \\
60671.9  & 11.3  & 385LP & G4                      & 3700--8800                     & \(\sim1300\)         & XL-216/BFOSC \\
60672.2  & 11.6  &       & Grism~\#4               & 3200--9600                     & \(\sim710\)         & NOT/ALFOSC \\
60672.9  & 12.3  & 385LP & G4                      & 3700--8800                     & \(\sim1300\)         & XL-216/BFOSC \\
60674.0  & 13.4  &       & Gr. 7, Gr. 8            & 3800--6840, 5800--8350         &\(\sim1300,\sim2200\)      & HCT/HFOSC \\
60674.9  & 14.3  & 385LP & G4                      & 3700--8800                     & \(\sim1300\)         & XL-216/BFOSC \\
60675.7  & 15.1  & 385LP & G4                      & 3700--8800                     & \(\sim1300\)         & XL-216/BFOSC \\
60676.9  & 16.3  & 385LP & G4                      & 3700--8800                     & \(\sim1300\)         & XL-216/BFOSC \\
60678.9  & 18.3  & 385LP & G4                      & 3700--8800                     & \(\sim1300\)         & XL-216/BFOSC \\
60679.8  & 19.3  & 385LP & G4                      & 3700--8800                     & \(\sim1300\)         & XL-216/BFOSC \\
60680.7  & 20.2  & 385LP & G4                      & 3700--8800                     & \(\sim1300\)         & XL-216/BFOSC \\
60681.9  & 21.3  & 385LP & G4                      & 3700--8800                     & \(\sim1300\)         & XL-216/BFOSC \\
60682.8  & 22.3  & 385LP & G4                      & 3700--8800                     & \(\sim1300\)         & XL-216/BFOSC \\
60683.9  & 23.3  & 385LP & G4                      & 3700--8800                     & \(\sim1300\)         & XL-216/BFOSC \\
60689.8  & 29.3  & 385LP & G4                      & 3700--8800                     & \(\sim1300\)         & XL-216/BFOSC \\
60690.8  & 30.3  & 385LP & G4                      & 3700--8800                     & \(\sim1300\)         & XL-216/BFOSC \\
60695.0  & 34.4  &       & 676R -- 420 gr/mm             & 3500--8950          &  \(\sim1160\) & DOT/ADFOSC \\
60696.9  & 36.3  & 385LP & G4                      & 3700--8800                     & \(\sim1300\)         & XL-216/BFOSC \\
60700.1  & 39.6  &       & Grism~\#4               & 3200--9600                     & \(\sim710\)         & NOT/ALFOSC \\
60703.1  & 42.5  &       & 676R -- 420 gr/mm            & 3500--8950         & \(\sim1160\)  & DOT/ADFOSC \\
60706.0  & 45.4  &       & 676R -- 420 gr/mm            & 3500--8950         &  \(\sim1160\)  & DOT/ADFOSC \\
60713.2  & 52.7  &       & 676R -- 420 gr/mm            & 3500--8950      &  \(\sim1160\)  & DOT/ADFOSC \\
60714.7  & 54.2  & 385LP & G4                      & 3700--8800                     & \(\sim1300\)         & XL-216/BFOSC \\
60732.0  & 71.5  &       & Grism~\#4               & 3200--9600                     & \(\sim710\)         & NOT/ALFOSC \\
60784.0  & 123.5 &       & R1000R                  & 5100--10000                    & \(\sim1000\)        & GTC/OSIRIS+ \\
60818.9  & 158.4 &       & R1000R                  & 5100--10000                    & \(\sim1000\)        & GTC/OSIRIS+ \\
\enddata

\tablecomments{(a) Epoch in days relative to discovery; (b) \citet{TNSCR4943}; (c) \citet{TNSCR4947}; (d) \citet{TNSCR4974}. (e) Resolving power \(R\!\equiv\!\lambda/\Delta\lambda\), listed as typical values.}
\end{deluxetable}

We obtained optical spectroscopy of SN~2024aecx with a number of 2–10\,m‐class telescopes. And these spectra are available online. At Xinglong Observatory, the 2.16 m reflector (XL-216) was equipped with the Beijing Faint Object Spectrograph and Camera (BFOSC), a Cassegrain spectrograph providing low-resolution spectra over the wavelength range from 3500–9000 \AA\ \citep{Fan2016}. The 2.0\,m Himalayan Chandra Telescope (HCT) at Hanle, outfitted with Hanle Faint Object Spectrograph Camera (HFOSC), covered 3500–9000\,\AA\ \citep{Cowsik2002,Prabhu2014}. India’s 3.6\,m Devasthal Optical Telescope (DOT) employed ADFOSC for low resolution spectra from 3500–10000\,\AA\ \citep{Sagar2020}. On La Palma, the 2.56\,m Nordic Optical Telescope (NOT) used ALFOSC and multiple grisms to span 3200–9600\,\AA\ \citep{Djupvik2010}. The 10.4\,m Gran Telescopio Canarias (GTC) at Roque de los Muchachos, equipped with OSIRIS, provided low and medium resolution spectra over the optical range \citep{Cepa2000}.

The first spectrum of SN\,2024aecx was obtained by XL-216 as early as $t=0.3$\,d after discovery, and we took another early spectrum with the GTC at $t=0.7$\,d. After that, our spectroscopic follow-up used a cadence of one to a few days out to $t=54.2$\,d. At the nebular phase, we took three more spectra with NOT and GTC at $t=71.5, 123.5$ and $158.4$\,d. The one-dimensional spectra were extracted with the data reduction packages \textsc{iraf} \citep{Tody1986} and \textsc{pypeit} \citep{Prochaska2020}. The raw spectral images were first bias-subtracted and flat-field corrected, followed by cosmic-ray removal and wavelength calibration using comparison arc lamps. Sky background was subtracted from regions adjacent to the target trace, and the spectra were flux-calibrated with spectrophotometric standard stars observed on the same nights.

Additionally, we also used three publicly available spectra from the TNS: two epochs obtained with Gemini North’s GMOS-N on the 8.1\,m telescope and one from the University of Hawaii 88-inch telescope (UH88; \citealt{Tucker2022}) \citep{TNSCR4943,TNSCR4947,TNSCR4974}. Table~\ref{tab:spec} shows a log of all spectroscopic observations used in this work.

\subsection{HST/ACS and VLT/MUSE Archive Data}
\label{sec:hstmuse}

The host galaxy of SN\,2024aecx, NGC\,3521, has been observed by the Advanced Camera for Survey (ACS) on the Hubble Space Telescope (HST) and the Multi Unit Spectroscopic Explorer (MUSE) on the Very Large Telescope (VLT). The fields of view of the HST/ACS and VLT/MUSE observations, along with the explosion site of SN\,2024aecx, are shown in Figure~\ref{fig:host}.

The HST/ACS observations (Program ID: GO-17502; PI: D. Thilker) imaged the outskirt of NGC\,3521 with exposure times of 1610\,s in the F606W band and 2110\,s in the F814W band. The data can be found in MAST:\dataset[10.17909/9dfy-qz66]{http://dx.doi.org/10.17909/9dfy-qz66}. In this work, we shall make use of these images to determine the distance to the host galaxy using the tip of the red giant branch (TRGB) method (Section~\ref{sec:dist}). The data were retrieved from the Mikulski Archive for Space Telescopes \footnote{\url{https://archive.stsci.edu/}} and details of data reduction shall be described in Section~\ref{sec:dist}.

VLT/MUSE provides integral-field-unit (IFU) spectroscopy over a $1\times1$~arcmin$^2$ field, spanning 4650--9300\,\AA\ \citep{bacon2010muse}. This wavelength range includes the principal nebular emission lines---H$\alpha$, H$\beta$, [O\,\textsc{iii}] $\lambda\lambda4959,5007$, and [N\,\textsc{ii}] $\lambda\lambda6548,6583$---from which gas-phase metallicity can be derived using strong-line diagnostics \citep{Pagel1979,Edmunds1984}. Such capabilities make VLT/MUSE exceptionally suited to probing the chemical properties of SN host regions. For NGC\,3521, VLT/MUSE obtained a deep 4500\,s exposure (Program ID: 099.B-0242, 098.B-0551; PI: C.M. Carollo) centered on the nucleus, and the field of view encompasses the SN explosion site. The VLT/MUSE data were retrieved from the ESO Science Archive\footnote{\url{https://archive.eso.org/scienceportal/home}}. Details of data reduction shall be described in Section~\ref{sec:metallicity}.

\section{Host galaxy}
\label{sec:host}

\begin{figure}
    \centering
    \includegraphics[width=1\linewidth]{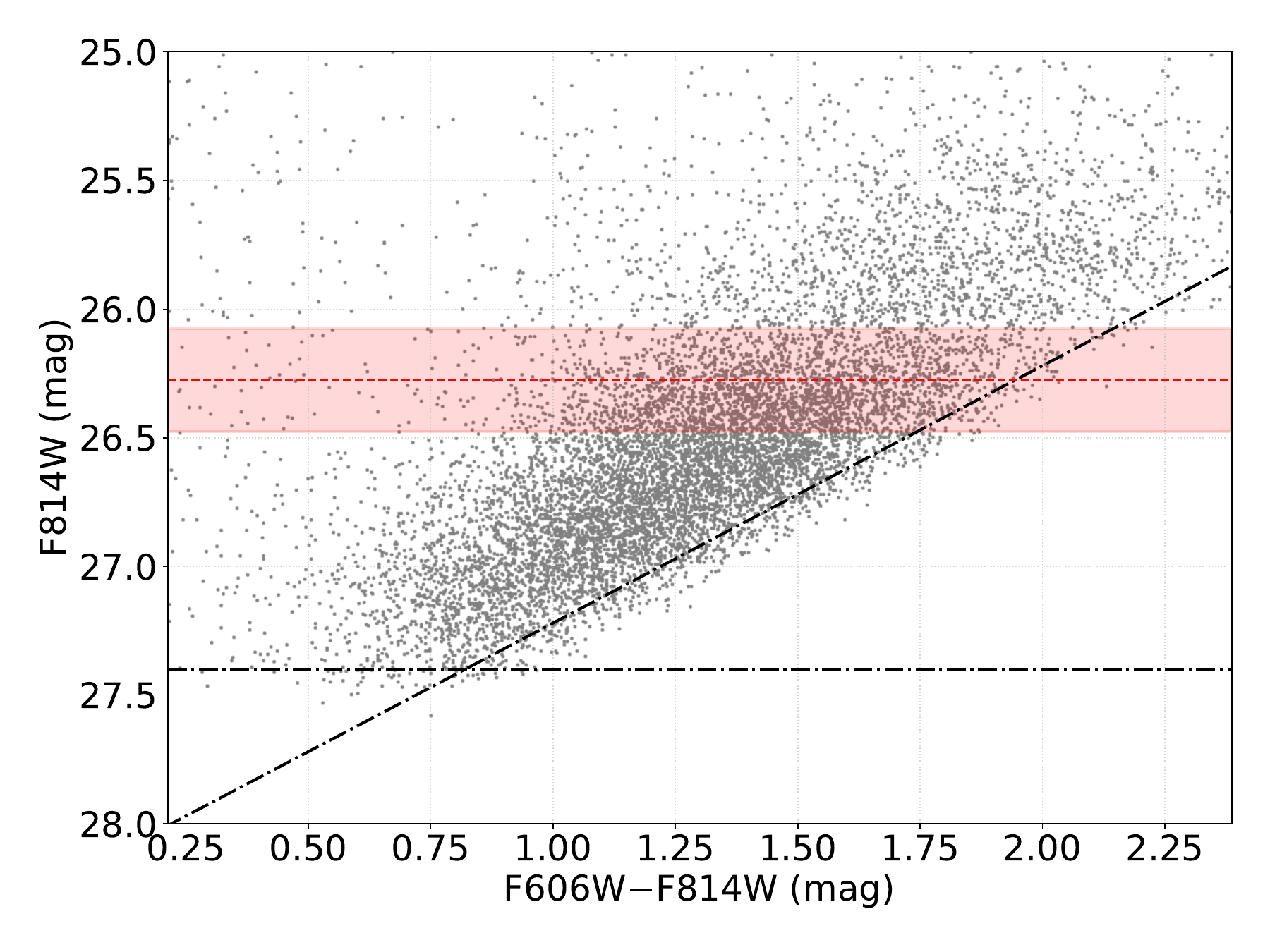}
    \includegraphics[width=1\linewidth]{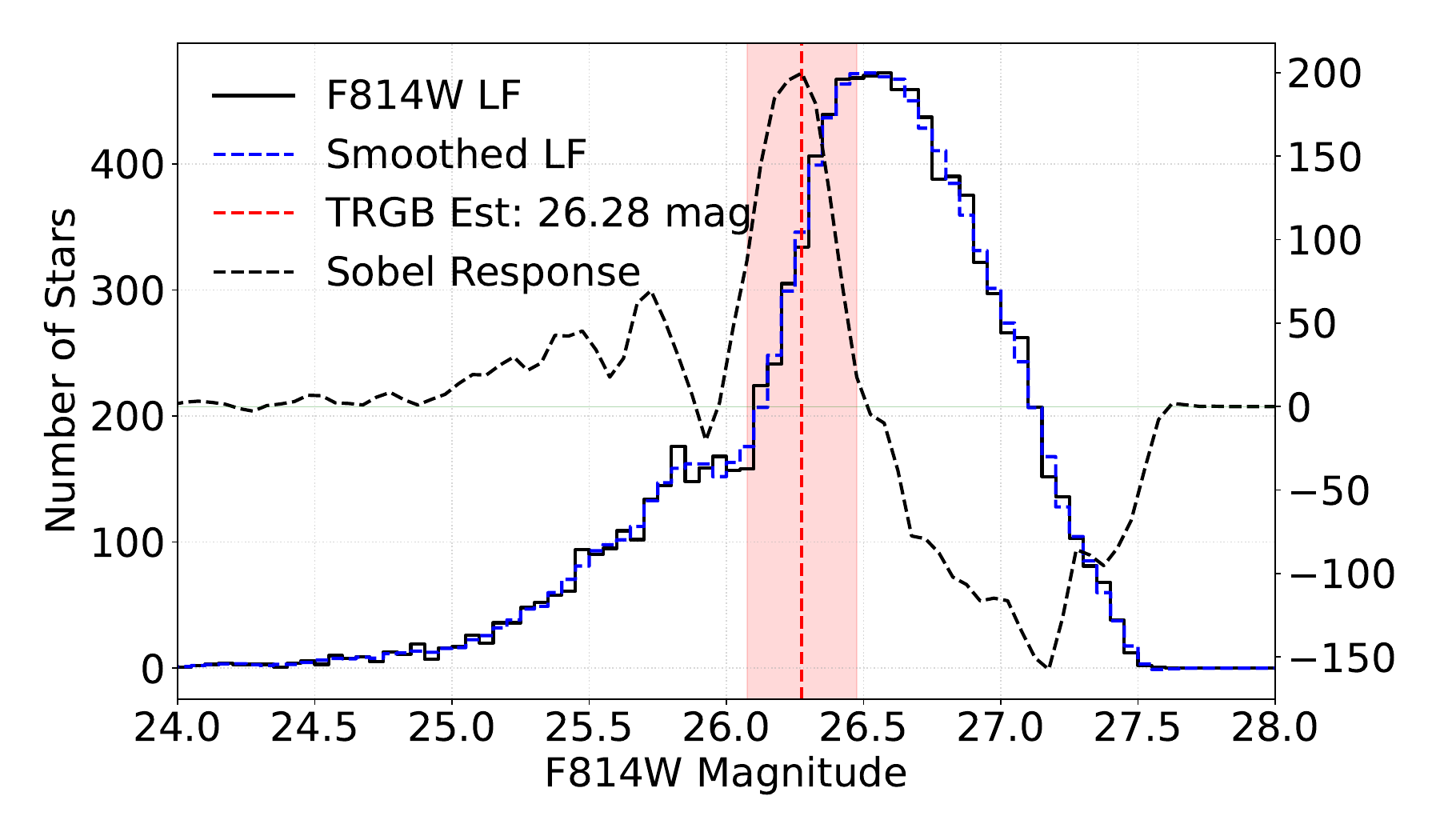}
    \caption{{\it Top:} Color–magnitude diagram of resolved stars in the outer regions of NGC\,3521 from {\it HST}/ACS imaging. Each gray point represents an individual star, plotted in F814W versus F606W–F814W. The red dashed line marks the measured TRGB at F814W$_0 = 26.28$\,mag, and the shaded band denotes its $\pm1\sigma$ uncertainty. The black dash–dot line indicates the approximate 50\% completeness limit. {\it Bottom:} F814W luminosity function of the observed stars (solid black histogram) together with its smoothed version (blue dashed line). The red vertical dashed line and shaded region again show the TRGB and its uncertainty. Overlaid in black (right-hand axis) is the Sobel edge–detection response, whose peak identifies the TRGB.}
    \label{fig:trgb}
\end{figure}

\begin{figure}
    \centering
    \includegraphics[width=1\linewidth]{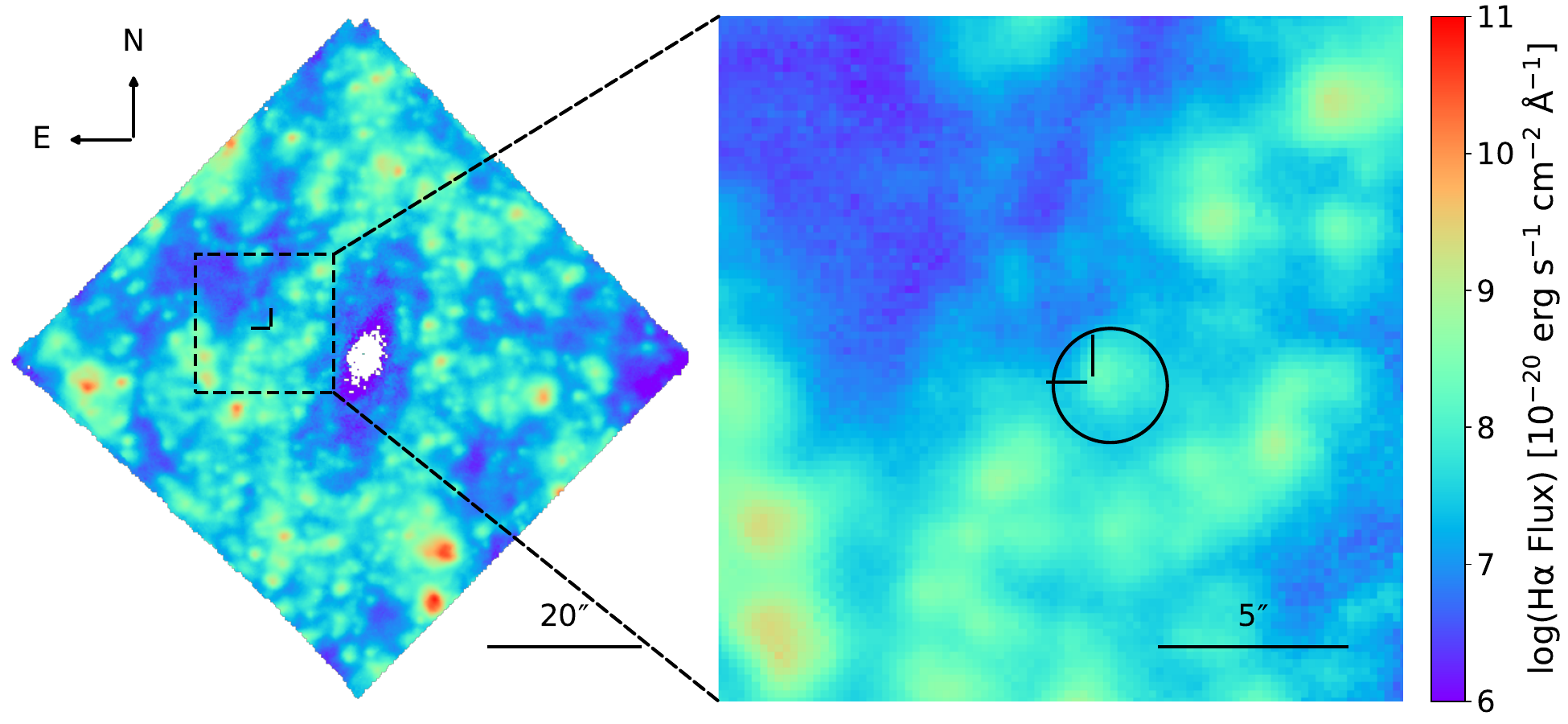}
    \includegraphics[width=1\linewidth]{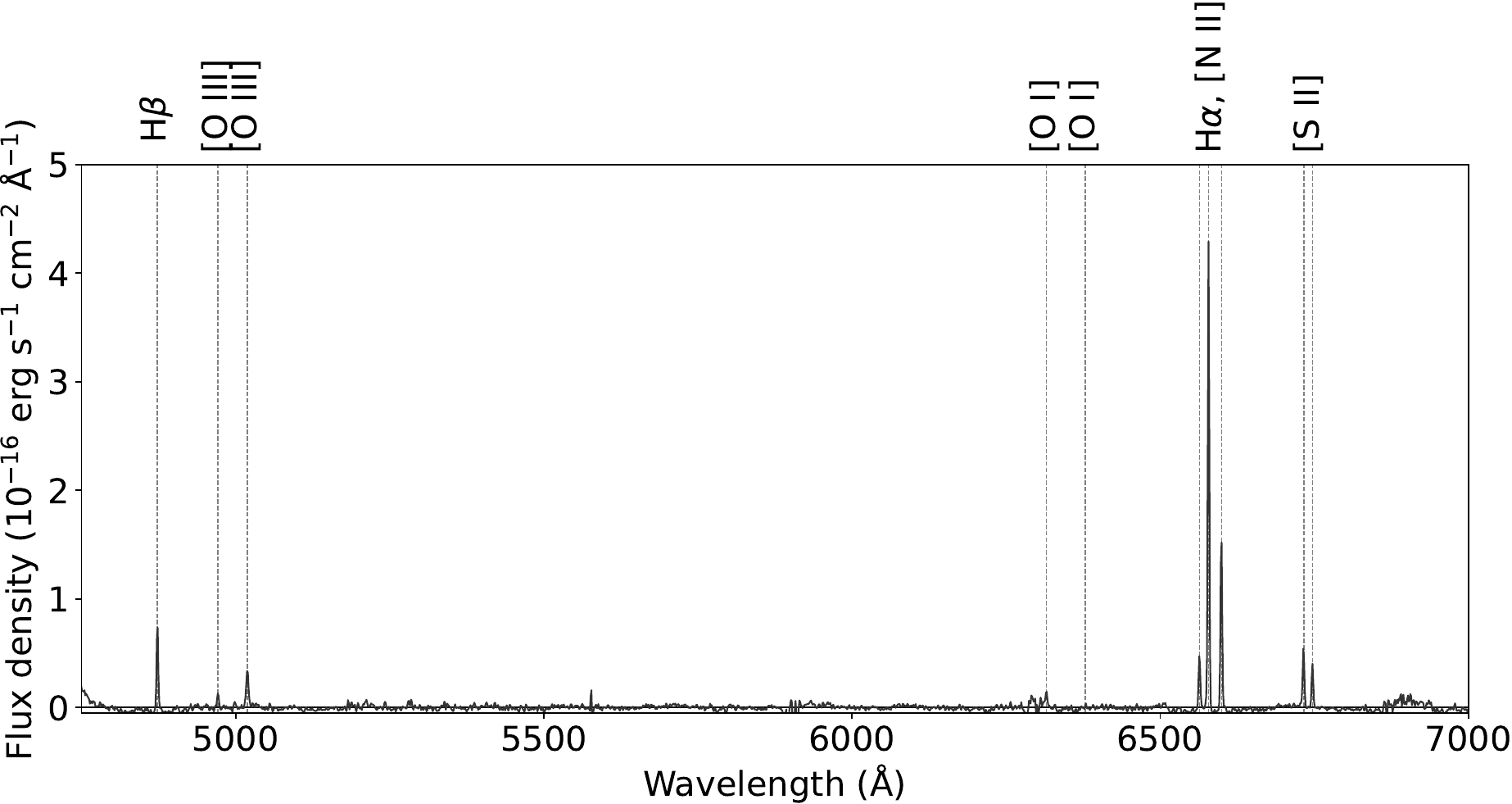}
    \caption{\textit{Top:} VLT/MUSE H$\alpha$ flux map of the environment of SN\,2024aecx in NGC\,3521, calculated from the 0-th moment of the datacube and shown on a logarithmic scale. The black cross mark the SN position and the circle outlines the local H\,II region, within which we extracted a spectrum for analysis (see text). North is up and east is left. \textit{Bottom:} Continuum‐subtracted spectrum extracted from the aperture.}
    \label{fig:muse}
\end{figure}

\begin{figure}
    \centering
    \includegraphics[width=1\linewidth]{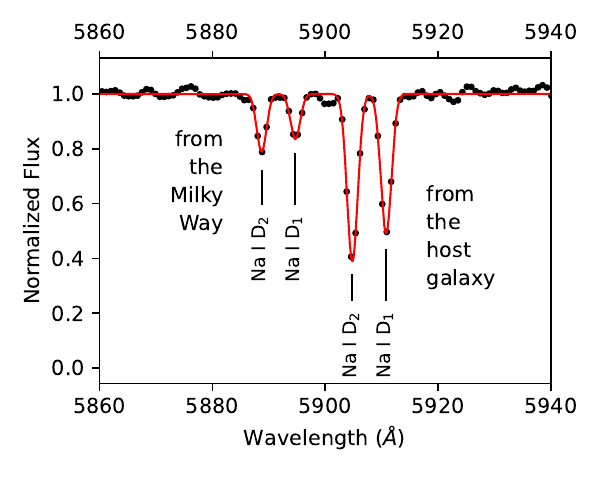}
    \caption{Normalized GTC spectrum of SN\,2024aecx obtained at $t=0.7$\,d, showing prominent Na\,\textsc{i}\,D absorption features. Black circles are the observed data, and the red curves are Gaussian fits to each line. The left component arises from the Milky Way, while the right component is from the host galaxy.}
    \label{fig:NaD}
\end{figure}

SN\,2024aecx is located in the central region of the nearby galaxy NGC\,3521, a late-type spiral galaxy that is closely analogous to the Milky Way (Fig.~\ref{fig:host}). In this section we try to constrain the distance, metallicity and extinction of the host galaxy, which are important for the analysis of SN\,2024aecx.


\subsection{Distance}
\label{sec:dist}
There are 26 redshift-independent distance measurements for NGC\,3521, all based on the Tully-Fisher method, according to the NASA/IPAC Extragalactic Database (NED)\footnote{\url{https://ned.ipac.caltech.edu/}}. The reported distances, however, span a very large range from 4.0\,Mpc to 16.4\,Mpc. In this work, we use the HST/ACS imaging observations of NGC\,3521 (Fig.~\ref{fig:host}; Section~\ref{sec:hstmuse}) and apply the TRGB method \citep{Baade1944,Sandage1971,Costa1990,Lee1993} to secure a more accurate measurement. This method is based on the fact that low-mass stars climbing the RGB reach a nearly fixed luminosity just before core helium ignition. Therefore, one can derive a precise distance by identifying the sharp cutoff in the observed luminosity function of RGB stars (i.e. the tip) and comparing its apparent magnitude with the well-calibrated TRGB absolute magnitude. This technique is relatively insensitive to metallicity and crowding effects, allowing for robust distance measurements for nearby galaxies.

We followed the same approach as described by \citet{Anand2020} to implement the TRGB distance measurement. We first performed PSF photometry on the HST/ACS images with the \textsc{dolphot} \citep{DOLPHOT} package, and then applied several quality cuts to construct a clean sample of stellar sources:
\begin{align}
\texttt{snr}_{\mathrm F606W} &> 5, \\[6pt]
\texttt{snr}_{\mathrm F814W} &> 5, \\[6pt]
\texttt{crowd}_{\mathrm F606W} + \texttt{crowd}_{\mathrm F814W} &< 0.8, \\[6pt]
\bigl(\texttt{sharp}_{\mathrm F606W} + \texttt{sharp}_{\mathrm F814W}\bigr)^2 &< 0.075,
\end{align}
where \texttt{snr}, \texttt{crowd} and \texttt{sharp} are \textsc{dolphot}-reported quality parameters on the signal-to-noise ratio, crowding and sharpness of the detected sources. The color-magnitude diagram of the stellar sample is shown in Fig.~\ref{fig:trgb} (top), where the RGB stars from NGC\,3521 are clearly visible.

We carried out artificial‐star tests in both the F606W and F814W bands to assess the photometry accuracy and detection limit. Different input magnitudes were considered, ranging from 25.0 to 28.7\,mag (25.0 to 28.0\,mag) for the F606W (F814W) band. For each magnitude, 6,000 artificial stars were randomly injected across the field and we tried to recover the artificial stars with \textsc{dolphot}. The probability of recovering sources above the 5$\sigma$ threshold varies with the input magnitude, which can be fitted with a Gaussian error function. We defined the detection limit as the magnitude where the probability declines to 50\%. The recovered magnitudes exhibit no systematic bias from the input magnitudes, but the scatters are systematically (but only slightly) larger than the nominal photometric errors reported by \textsc{dolphot}. We applied a scaling factor to correct the photometric errors in our stellar sample.

We binned the F814W magnitudes in intervals of 0.05\,mag to construct the luminosity function, which was then smoothed using a Savitzky–Golay filter (window length of 7 bins and a quadratic polynomial fit) \citep{Savitzky1964} to suppress shot noise while preserving the intrinsic discontinuity at the tip (Fig.~\ref{fig:trgb}, bottom). A Sobel edge‐detection filter ([–2, 0, +2]) was applied to the smoothed luminosity function \citep{Lee1993}. We regard the peak of the Sobel response as the TRGB magnitude and the half‐width at half‐maximum as the uncertainty, which is $26.28 \pm 0.21$\,mag. With this value, we derived a distance of $11.3 \pm 1.1$\,Mpc for NGC\,3521, adopting the TRGB zero point (i.e., color calibration) of \citet{Rizzi2007} and incorporating systematic uncertainties associated with the zero-point as described in \citet{Anand2020}. We have considered the Galactic extinction \citep{SFD1998}; internal extinction within the host galaxy is negligible for the analyzed HST/ACS field, which lies in the far galaxy outskirt. And this distance is consistent, within the uncertainties, with the value of $11.40 \pm 0.56$\,Mpc adopted by \citet{Zou2025}.

\subsection{Metallicity}
\label{sec:metallicity}

In this section we try to derive the metallicity of SN\,2024aecx based on VLT/MUSE IFU spectroscopy. We processed the VLT/MUSE datacube with the \textsc{ifuanal} package \citep{Lyman2018}, following the same methodology of \citet{Sun20222019yvr, Sun2023} and \citet{Xi2025}. First, the datacube was de-redshifted and corrected for Galactic extinction. The spaxels were then adaptively binned using a Voronoi tessellation scheme until achieving a continuum signal-to-noise ratio of at least 120 over the 5590–-5680\,\AA\ interval. In each Voronoi bin, the combined spectrum was fitted for its stellar continuum with the \textsc{starlight} package \citep{Fernandes2005}, employing \citet{Bruzual2003} simple stellar population models. Our library comprised 60 base models spanning 15 ages (3 Myr–13 Gyr) and 4 metallicities (Z = 0.004, 0.008, 0.02, 0.05). The best-fit stellar continuum in each bin was then scaled appropriately and subtracted from the observed spectra at the individual spaxel level.

Figure~\ref{fig:muse} (top) shows a map of H$\alpha$ integrated flux, which is calculated with the 0-th momentum. The observed field is full of ionized gas, suggesting very active star formation. The position of SN\,2024aecx coincides with a circular H\textsc{ii} region with a radius of $R = 80$\,pc. We co-added spectra from all spaxel within this region and the stacked spectrum is displayed in the bottom panel of Fig.~\ref{fig:muse}. The spectrum exhibits the strong Balmer lines of H\(\alpha\) (\(\lambda6563\)) and H\(\beta\) (\(\lambda4861\)) as well as prominent forbidden transitions of [O\,\textsc{iii}] \(\lambda\lambda4959,5007\), [N\,\textsc{ii}] \(\lambda\lambda6548,6583\), and [S\,\textsc{ii}] \(\lambda\lambda6716,6731\). In contrast, the lines of [O\,\textsc{i}] \(\lambda\lambda6300\,6363\) are very faint. We fitted each line with a Gaussian function to measure its integrated flux.

We estimated the gas‐phase metallicity with the strong‐line method based on the O3N2 calibration of \citet{Marino2013}:
\[
12 + \log(\mathrm{O/H}) = 8.533 - 0.214 \times \mathrm{O3N2},
\]
where
\[
\mathrm{O3N2} \;=\; \log\!\biggl(\frac{[\mathrm{O\,\textsc{iii}}]\lambda5007}{\mathrm{H}\beta}\biggr) \;-\; \log\!\biggl(\frac{[\mathrm{N\,\textsc{ii}}]\lambda6584}{\mathrm{H}\alpha}\biggr).
\]
Applying this calibration to our measured line ratios, we derived an oxygen abundance of $12 + \log(\mathrm{O/H}) = 8.49 \pm 0.18$\,dex, which is very close to the solar value (8.69\,dex, \citealt{Asplund2009}).

\subsection{Extinction}
\label{sec:extinction}

The Galactic extinction toward NGC\,3521 is $A_V^{\rm MW} = 0.144$\,mag according to the \cite{SFD1998} all-sky dust map. We employed three different methodologies to derive the internal extinction of SN\,2024aecx due to interstellar dust within the host galaxy. For simplicity, we assume a \citet{Cardelli1989} standard extinction law with $R_V = 3.1$.

\subsubsection{Balmer Decrement}

As described in Section~\ref{sec:metallicity}, SN\,2024aecx is spatially coincident with an H\textsc{ii} region. Its VLT/MUSE spectrum (Figure~\ref{fig:muse}) allows us to estimate an extinction with the Balmer‐decrement method. The observed flux ratio of H$\alpha$/H$\beta$ is 5.33 while the intrinsic flux ratio should be close to 2.87 \citep{Osterbrock2006}. From this we derived an extinction of $A_V^{\rm Host} = 1.95$\,mag. It should be noted, however, that the Balmer-decrement extinction corresponds to that for the ionized gas and could be different from that for the SN if there is dust between them along the line of sight \citep[e.g.][]{Sun20212004dg}.

\subsubsection{Na\,\textsc{i}\,D Absorption}

At $t=0.7$\,d, we obtained a deep spectrum with the GTC telescope (see Sections~\ref{sec:obs} and \ref{sec:spectroscopy}), where the Na\,\textsc{i}\,D absorption features are clearly resolved. As shown in Fig.~\ref{fig:NaD}, there are two components of Na\,\textsc{i}\,D absorption, one close to the doublet's rest wavelengths and one at a higher redshift of $z$ = 0.0025 that is consistent with NGC\,3521. We attribute the former to interstellar absorption within the Milky Way and the later to that in the host galaxy. We modeled the continuum as a constant and fitted each line with a Gaussian profile. The total equivalent width (EW) of the host-galaxy Na\,\textsc{i}\,D doublet is ${\rm EW^{Host}} = (1.21\pm0.03$)\,\AA\,$+ (1.45\pm0.03)$\,\AA\,$= 2.65\pm0.04$\,\AA, consistent with the measurement of \citet{Zou2025}. 

A number of works reported empirical relations between the Na\,\textsc{i}\,D absorption and interstellar extinction/reddening \citep[e.g.][]{Turatto2002, Poznanski2012, Stritzinger2018}. The host-galaxy Na\,\textsc{i}\,D EW is outside the dynamical range of the relation of \citet{Poznanski2012} (which has a maximum value of EW\,=\,2.4\,\AA) and applying their relation will lead to an unphysically large value of extinction. We can derive an extinction of $A_V^{\rm Host} = 2.1$~mag based on \citet{Stritzinger2018}; however, there is significant scatter in their relation and the derived value could have a large uncertainty. \citet{Turatto2002} found two branches in the relation between reddening and Na\,\textsc{i}\,D EW (their Fig.~3). Adopting the lower branch we infer a host extinction of $A_V^{\rm Host} = 1.3$\,mag for SN\,2024aecx. 

\cite{Rodriguez2023} presented an empirical relation between the host-galaxy extinction of SESNe and the equivalent width of the Na\,\textsc{i}\,D absorption feature, \(E(B{-}V)=(0.007\pm0.024) + (0.246\pm0.054)\,\mathrm{EW}_{\mathrm{Na\,ID}}\)\,{[\AA]}, with an intrinsic scatter \(\hat{\sigma}=0.094\)~mag. Substituting \(\mathrm{EW}_{\mathrm{Na\,ID}}=2.7\)~\AA gives \(E(B{-}V)=0.007 + 0.246\times 2.7 = 0.67\)~mag. Propagating the coefficient uncertainties together with the intrinsic scatter yields \(\sigma_{E(B{-}V)}\approx 0.18\)~mag, i.e., \(E(B{-}V)=0.67\pm0.18\)~mag. However, the fit exhibits substantial intrinsic dispersion, and the calibration sample in \cite{Rodriguez2023} covered only \(\mathrm{EW}_{\mathrm{Na\,ID}}\approx 0-1\)~\AA, well below our measurement; consequently, this extrapolated estimate may not be reliable.


\subsubsection{Comparison with template color}

One can also derive extinction by comparing the observed SN color curves with well-established templates \citep{Stritzinger2018}. With this method we obtained $A_V^{\rm Host} = 1.2$\,mag, the details for which shall be described later in Section~\ref{sec:color}. This value is very close to the result based on Na\,\textsc{i}\,D absorption and is adopted for subsequent analysis. The value is much smaller than that based on Balmer decrement, suggesting that there is a non-negligible amount of dust along the line of sight between SN\,2024aecx in the foreground and the ionized gas in the background.

\section{Light curve}
\label{sec:lightcurve}

\begin{figure*}
    \centering
    \includegraphics[width=1\linewidth]{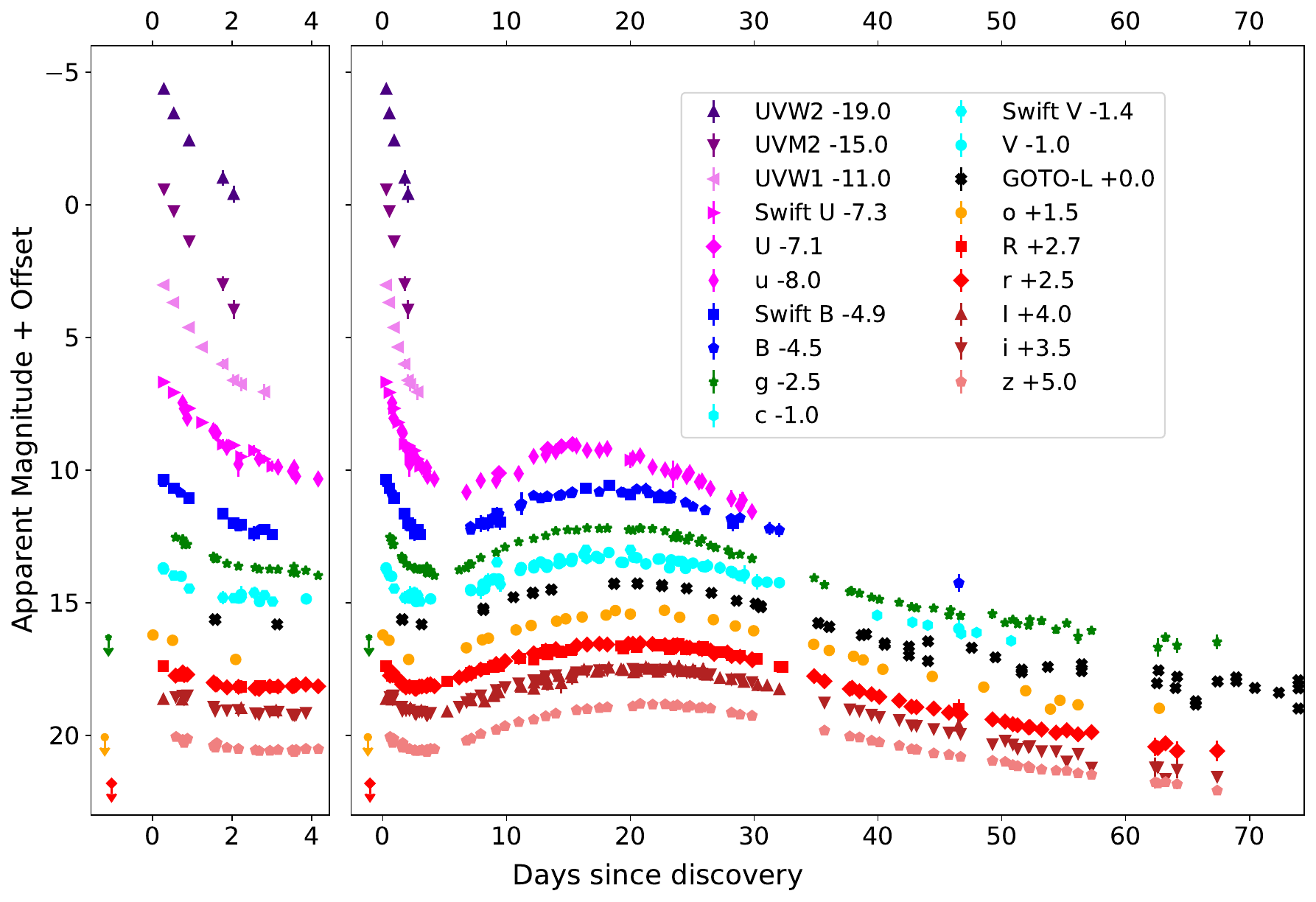}
    \caption{Multi-band light curves of SN\,2024aecx. Each filter is shifted vertically for clarity. Error bars denote $1\sigma$ photometric uncertainties.}
    \label{fig:LC}
\end{figure*}

\begin{figure}
    \centering
    \includegraphics[width=1\linewidth]{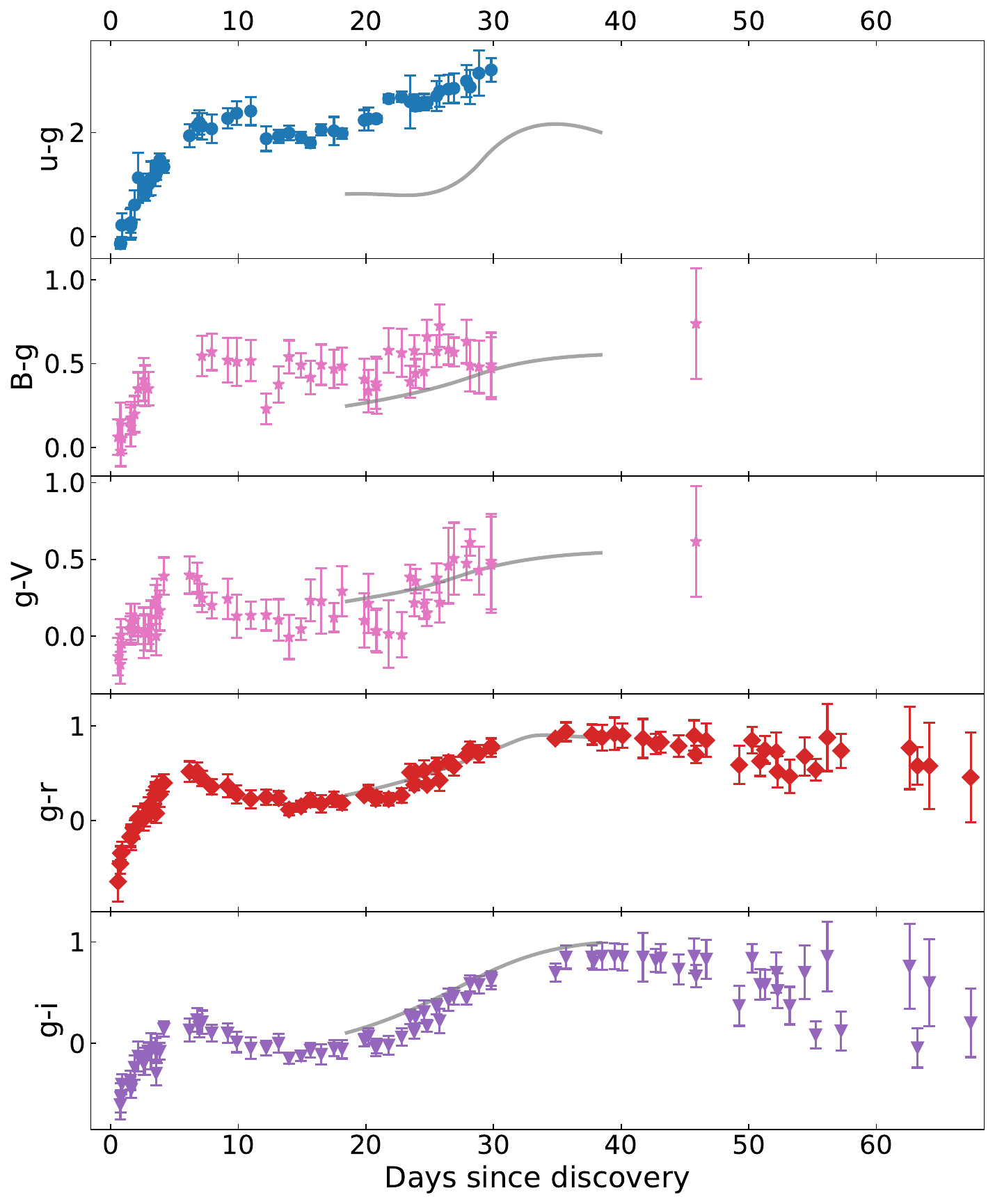}
    \caption{Color evolution of SN\,2024aecx after correcting for the Galactic and host extinctions ($A^{\mathrm{tot}}_V = 1.34$). The orange lines are template color curves for Type~IIb SNe \citep{Stritzinger2018}.}
    \label{fig:color}
\end{figure}

\begin{figure}
    \centering
    \includegraphics[width=1\linewidth]{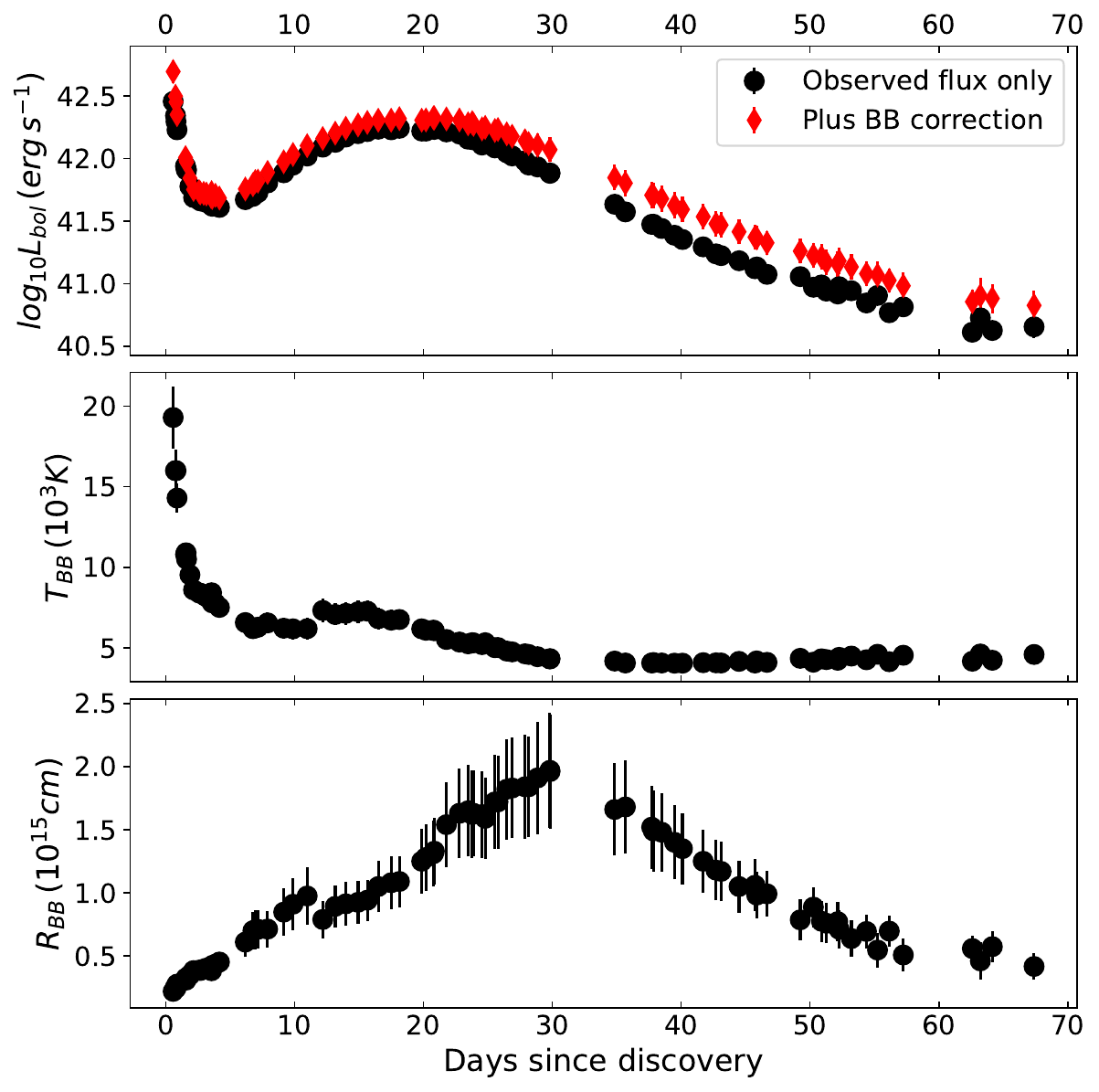}
    \caption{%
    Bolometric evolution of SN\,2024aecx derived with the \textsc{superbol} code. \textit{Top:} Logarithmic bolometric luminosity $L_{\rm bol}$ from the observed flux only (black circles) and after applying the blackbody correction (red diamonds). \textit{Middle:} Blackbody temperature $T_{\rm BB}$ in units of $10^3\,$K. \textit{Bottom:} Blackbody radius $R_{\rm BB}$ in units of $10^{15}\,$cm. All quantities are plotted as a function of days since discovery.%
  }
    \label{fig:bol}
\end{figure}

Figure~\ref{fig:LC} presents the UV–optical multi-band light curves of SN\,2024aecx. The last non-detection by ATLAS is at $t=-1.2$\,d with a detection limit of $o>18.56$\,mag while the last non-detections by ZTF are at $t=-1.1$\,d with $g>18.8$\,mag and $t=-1.0$\,d with $r>19.3$\,mag. This suggests that SN\,2024aecx was discovered shortly after its explosion, likely within 1\,day. At $t\sim0$\,day, SN\,2024aecx appears to be very bright and blue, with strong emission in the UV filters. During the next few days, the flux decreases swiftly across all bands, with steeper declines in the bluer filters and more gradual fading at redder wavelengths. Such behavior is characteristic of the shock-cooling emission observed in Type~IIb SNe, as exemplified by SN\,1993J, SN\,2011dh, and SN\,2016gkg \citep{Schmidt1993J,Ergon20142011dh,Tartaglia2016gkg}.

Following the shock‐cooling phase, SN\,2024aecx falls below the detection limit in the UV filters (Swift/UVOT $W2$, $M2$ and $W1$) while in the optical bands it rises again to a broad second maximum. The $u$ band peaks at \(t \simeq 15\)\,days, while the $r$, $i$, and $z$ bands reach their maxima later at \(t \simeq 20\) days. The systematically longer rise times at redder wavelengths reflect the increased photon diffusion time through the cooler, more opaque ejecta \citep{Taddia2018}. After the main peak, the light curves gradually decline out to 70~days post discovery.

\subsection{Color evolution}
\label{sec:color}

Figure~\ref{fig:color} shows the $X-g$ or $g-X$ ($X=u, B, V, r, i$) color curves of SN\,2024aecx in comparison to the \citet{Stritzinger2018} template of Type\,IIb SNe, which covers 20\,days from the $V$-band maximum. In constructing the color curves, we have interpolated the epochs of photometry in each band with respect to those in the $g$ band.

The observed colors are much redder than the template even after correcting for a Galactic extinction of $A_V=0.144$\,mag (\citealt{SFD1998}; Section~\ref{sec:extinction}). This suggests that SN\,2024aecx suffers from significant interstellar dust reddening within the host galaxy. We fitted the color curves with the least-square method by varying the host extinction; under the assumption of a standard \citet{Cardelli1989} extinction law with $R_V=3.1$, the observed colors match the template best with a host extinction of $A_V^{\rm Host}=1.2$\,mag. This value is very consistent with that derived with Na\,\textsc{i}\,D absorption (1.3\,mag; see Section~\ref{sec:extinction}).

Note that, however, the $u-g$ color still appear much redder than the template after correcting for the Galactic and host extinction. We tested by changing $R_V$ in the extinction law and refitting the color curves; in all cases, no better match can be found. We speculate that this is due to some strong absorption (e.g. from the iron-group elements), which makes SN\,2024aecx very faint in the $u$ band.

During the shock cooling phase, SN\,2024aecx shows a very blue color ($g-r= -0.6 $\,mag) at $t\sim0$\,day, followed by a rapid blue-to-red evolution to $g-r= 0.5$\,mag at $t=4$\,d. After that, it again turns bluer until $g-r=0.1 $\,mag near maximum at $t=15$\,d, and then evolves redward after maximum. From $t=35\sim40$\,d, SN\,2024aecx starts to slowly evolve blueward. This color evolution is similar to those observed for other Type\,IIb SNe \citep{Ergon20142011dh,Schmidt1993J,Tartaglia20172016gkg,Stritzinger2018,Kumar2013,Morales-Garoffolo2014,Morales-Garoffolo2015,Subrayan2025}.

\subsection{Bolometric Light Curve}

We employed the \textsc{superbol} package \citep{Nicholl2018superbol} to reconstruct the UV-optical (pseudo‐)bolometric light curve. In doing this we have adopted a distance of $D=11.3$\,Mpc, a Galactic extinction of $A_V^{\rm MW}=0.144$\,mag, a host extinction of $A_V^{\rm Host}=1.2$\,mag, and a standard extinction law of \citet{Cardelli1989} with $R_V=3.1$ (see Section~\ref{sec:host}). The results are shown in Fig.~\ref{fig:bol}.

At $t=0$\,d, SN\,2024aecx has a very high bolometric luminosity of $\mathrm{log}(L_{\rm bol}/\mathrm{erg}\,\mathrm{s}^{-1})\gtrsim42.5$ and blackbody temperature of $T_{\rm BB}\sim20,000$\,K. At $t=4$\,d, the bolometric luminosity declines to a local minimum of $\mathrm{log}(L_{\rm bol}/\mathrm{erg}\,\mathrm{s}^{-1})\sim 41.7 $ while the blackbody temperature cools down to $T_{\rm BB}=7510$\,K. After the shock cooling phase, the bolometric luminosity climbs up to the main peak with $\mathrm{log}(L_{\rm bol}/\mathrm{erg}\,\mathrm{s}^{-1})\sim 42.2 $ at $t=20$\,d and then gradually decreases; the blackbody temperature reaches the second local maximum with $T_{\rm BB}=7300$\,K at $t=15$\,d, after which it cools down continuously. The blackbody radius follows an almost linear evolution with time before and after a peak of $R_{\rm BB}=2\times10^{15}$\,cm at $t=30$\,d. It should be noted that, after the main peak, a significant portion of the SN radiation may lie at infrared wavelengths outside our observed UV and optical filters. In constructing the bolometric light curve, the code fits a blackbody spectral energy distribution (SED) to the multi-band UV–optical fluxes at each epoch, and the NIR contribution is obtained by interpolating this best-fitting blackbody SED. In our case, the inferred NIR flux contributes about 30\% of the total flux in the late time, so its omission has only a minor effect on the derived explosion parameters.

\subsection{Comparison with Other SNe}
\begin{figure*}
    \centering
    \includegraphics[width=1\linewidth]{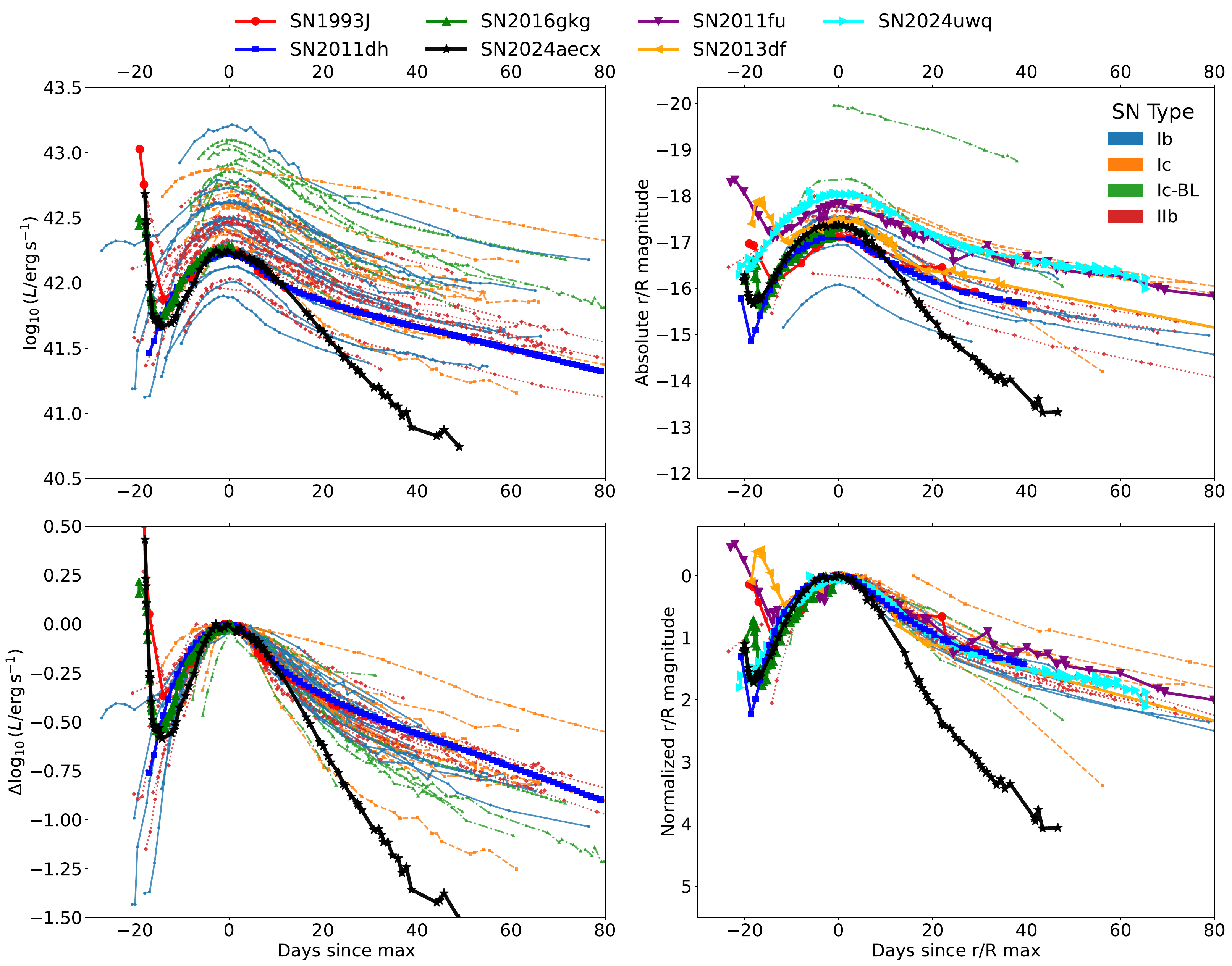}
    \caption{\textit{Left:} Bolometric light curves for SN\,2024aecx (black stars) and a sample of SESNe, with the top panel showing absolute luminosities and the bottom panel showing the same curves normalized to their peaks. 
\textit{Right:} $r$/$R$-band light curves for SN\,2024aecx (black stars) and the same SESNe sample, with the top panel showing absolute magnitudes and the bottom panel showing the normalized curves. All epochs are relative to days since maximum. The bolometric light curves are drawn from \citet{Lyman2016} and \citet{Taddia2018} while the $r$-band light curves are taken from \citet{Taddia2018}}
    \label{fig:comparison}
\end{figure*}

Figures~\ref{fig:comparison} compare the (pseudo-)bolometric light curve and the $r$/$R$-band light curve of SN\,2024aecx with those of other SESNe. In particular, we show three well-studied Type~IIb events SN\,1993J, SN\,2011dh and SN\,2016gkg \citep{Schmidt1993J,Arcavi2011,Tartaglia20172016gkg}, which are highlighted in the figures. For the other SESNe, the bolometric light curves are drawn from \citet{Lyman2016} and \citet{Taddia2018} while the $r$-band light curves are taken from \citet{Taddia2018}. To facilitate a direct comparison of their temporal behavior, we also present normalized light curves whose peak logarithmic luminosities (or peak magnitudes) are set to zero.

Before SN\,2024aecx, some Type\,IIb SNe are also found to have early peaks (e.g. SN\,1993J, SN\,2011dh, SN\,2011fu, SN\,2013df, SN\,2016gkg and SN2024uwq; \citealt{Schmidt1993J,Ergon20142011dh,Kumar2013,Morales-Garoffolo2014,Morales-Garoffolo2015,Tartaglia20172016gkg,Subrayan2025}), while other SNe of this class do not show obvious signs of early shock-cooling emission. Compared with them, SN\,2024aecx has one of the most prominent early shock-cooling peaks. The luminosity of the first peak [$\mathrm{log}(L_{\rm bol}/\mathrm{erg}\,\mathrm{s}^{-1})\gtrsim42.5$] and the change of luminosity till the dip ($\Delta\mathrm{log}(L_{\rm bol}/\mathrm{erg}\,\mathrm{s}^{-1})\sim1.0$\,dex) both have the largest values for Type\,IIb SNe. The duration of the first peak is also very short ($\sim$4\,d) compared with other Type\,IIb SNe.

The bolometric luminosity and $r$-band absolute magnitude of the main peak of SN\,2024aecx are relatively low but still within the observed ranges of Type\,IIb SNe. However, the rise to the main peak and the post-peak decline are markedly fast compared with other SESNe. This is most obvious in the normalized $r$-band light curve (Fig.~\ref{fig:comparison}, right), which shows that the $r$-band magnitude fades by approximately $\sim$4\,mag over just 40 days. The reason of this rapid evolution will be discussed further in Section~\ref{sec:model}.

\section{Spectroscopy}
\label{sec:spectroscopy}

\subsection{Photosphere phase spectrum}
\begin{figure*}
    \centering
    \includegraphics[width=1\linewidth]{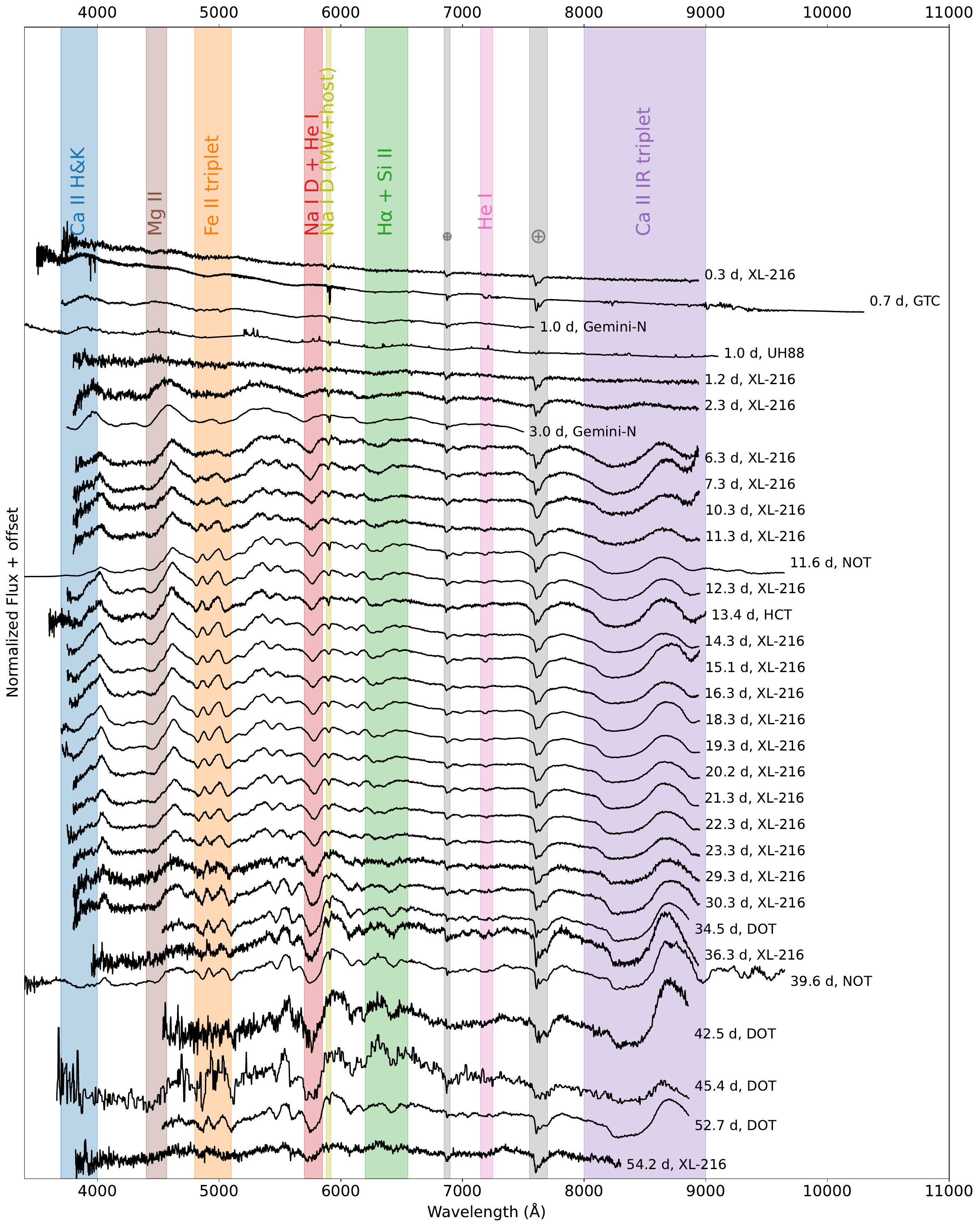}
    \caption{%
    Flux‐normalized spectra of SN\,2024aecx from $t=0.3$\,d to $t=54.2$\,d (uncorrected for redshift or extinction). Major telluric bands are shaded in gray and key SN features are indicated by colored bands.%
  }
    \label{fig:spec}
\end{figure*}

\begin{figure*}
    \includegraphics[width=1\linewidth]{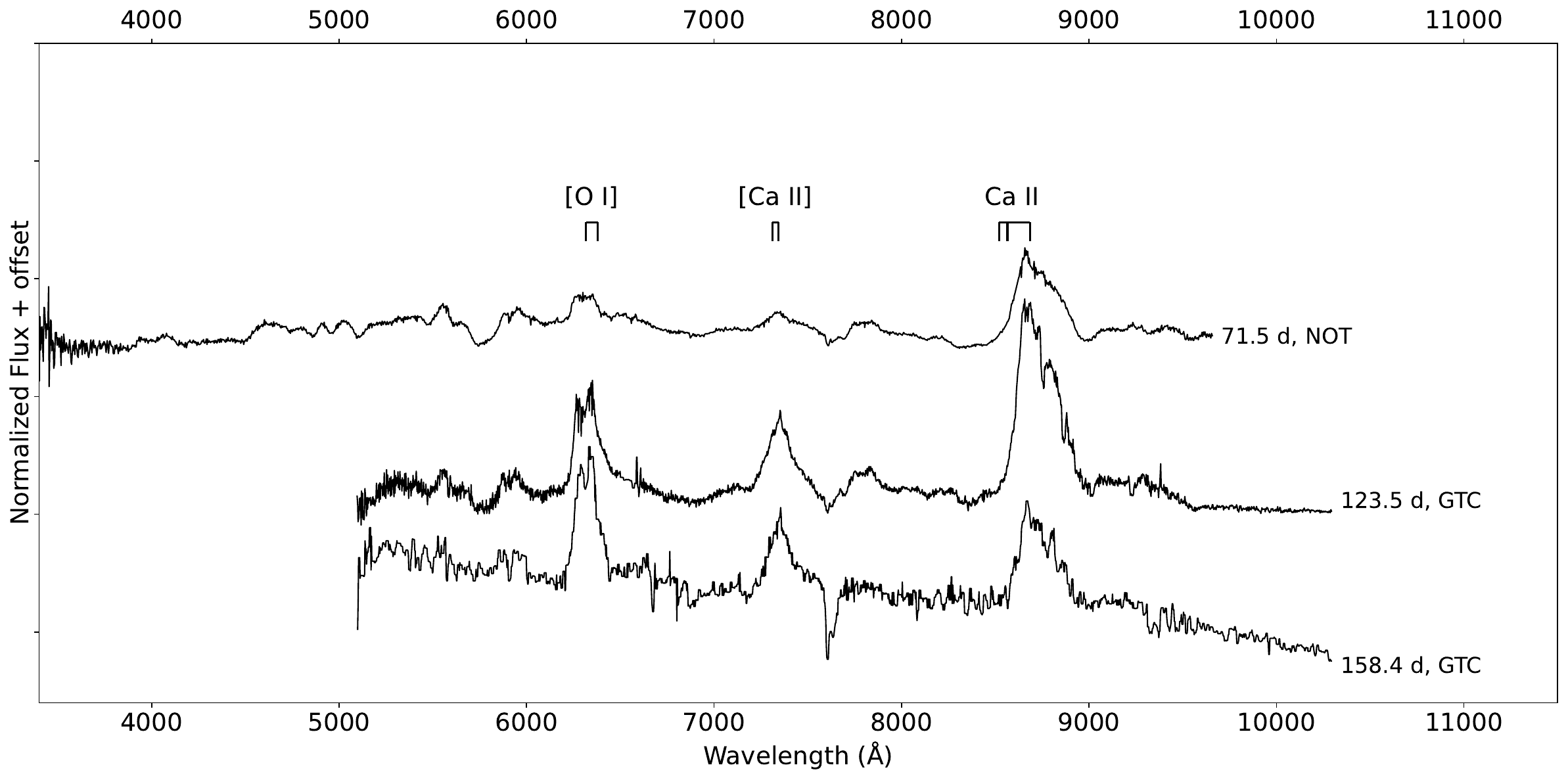}
\caption{
   Nebular-phase spectra of SN\,2024aecx at $71.5$\,d, $123.5$\,d, and $158.4$\,d (uncorrected for redshift or extinction), with the main emission lines marked.}
    \label{fig:nebula}
\end{figure*}

\begin{figure}
    \centering
    \includegraphics[width=1\linewidth]{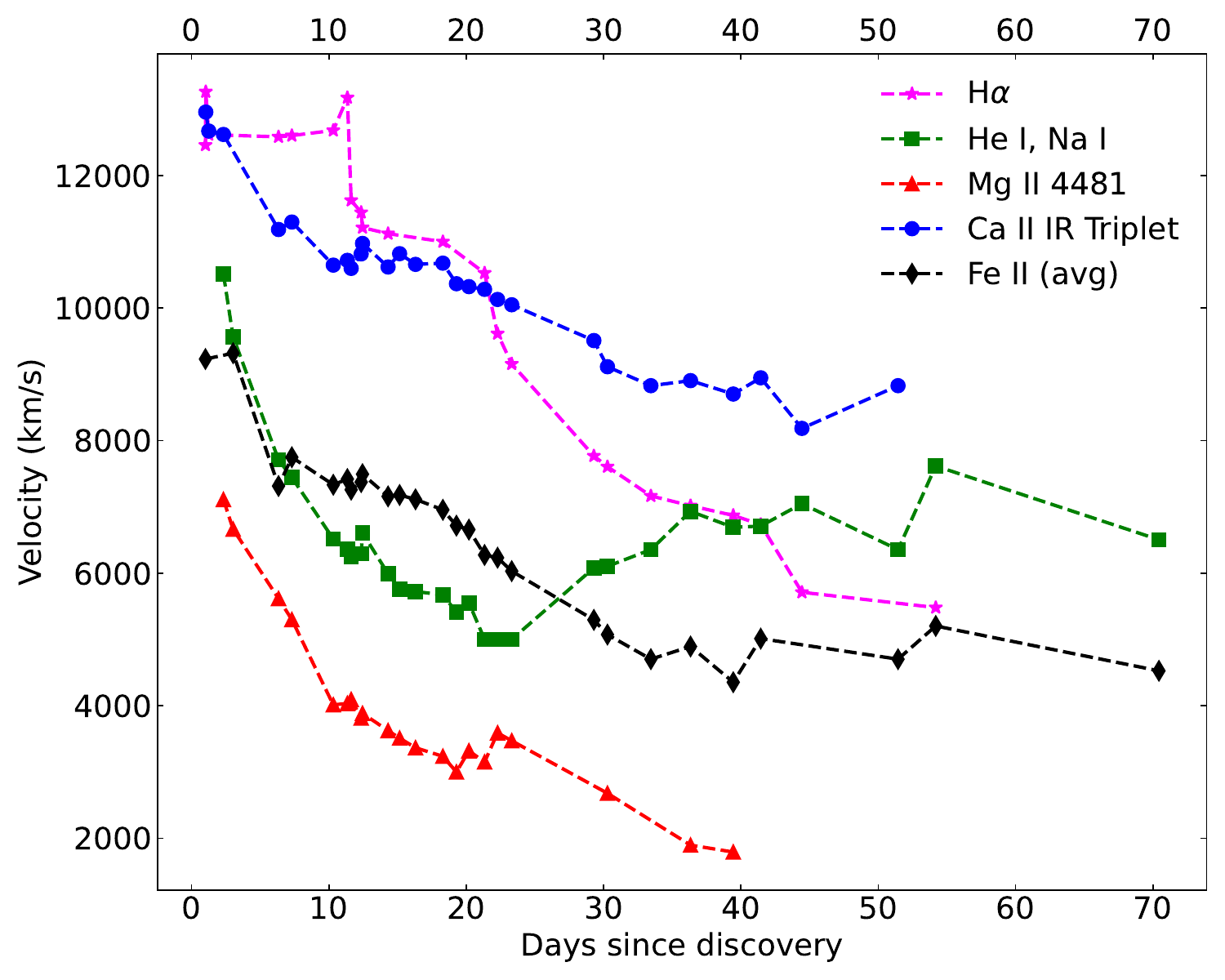}
  \caption{Evolution of line velocities for SN\,2024aecx measured from Gaussian fits to the absorption minima.}
    \label{fig:velocity_evolution}
\end{figure}

Figure~\ref{fig:spec} and~\ref{fig:nebula} shows the complete spectral sequence of SN~2024aecx (uncorrected for redshift or extinction), spanning from the earliest post–explosion epoch to the nebular phase. At the earliest epochs of $t=0.3-1.2$\,d, SN~2024aecx displays a hot, blue, and nearly featureless continuum with only Na\,\textsc{i}\,D absorptions. Its early-time appearance closely resembles other young SNe (e.g. SN\,2011dh and SN\,2016gkg; \citealt{Arcavi2011,Tartaglia2016gkg}). Using the Na\,\textsc{i}\,D absorption in SN~2024aecx, we infer the host-galaxy extinction as detailed in Section~\ref{sec:extinction}.

From $t=2.3$\,d, the SN spectra develop pronounced P‐Cygni profiles in Ca\,\textsc{ii} H\&K: $\lambda\lambda\,3933.66,\,3968.47\,\text{\AA}$, Ca\,\textsc{ii} IR triplet: $\lambda\lambda\,8498.02,\,8542.09,\,8662.14\,\text{\AA}$, the Mg\,\textsc{ii}\,$\lambda4481$, and the Fe\,\textsc{ii}\,$\lambda\lambda4924,5018,5169$ triplet. There is also a P-Cygni feature associated with He\,\textsc{i} $\lambda$5876, which is possibly blended with the Na\,\textsc{i}\,D lines. There are some weak features at the wavelength of H$\alpha$, which is, however, very easily confused with Si\,\textsc{ii}\,$\lambda6355$. \citet{Zou2025} used \textsc{synapps} synthesis and suggested that this feature arises primarily from both H$\alpha$ and Si\,\textsc{ii}. This confirms SN\,2024aecx to be a Type~IIb event.

We modeled the absorption minima of the P–Cygni profiles using Gaussian functions to derive the velocity evolution of the spectral lines (shown in Fig.~\ref{fig:velocity_evolution}). Initially, the velocities of all spectral lines decrease with time. The Ca\,\textsc{ii} infrared triplet and H\,$\alpha$ exhibits the highest velocity, reaching $\sim12{,}000\ \mathrm{km\,s^{-1}}$. This is followed by He\,\textsc{i}\,$\lambda5876$ and the Fe\,\textsc{ii} triplet, both peaking at nearly $10{,}000\ \mathrm{km\,s^{-1}}$. The Mg\,\textsc{ii}\,$\lambda4481$ line shows the lowest velocity, starting at approximately $7{,}000\ \mathrm{km\,s^{-1}}$ and declining thereafter.
The different velocities measured at the same epoch reflect the stratification of the ejecta: strong lines such as H$\alpha$ and Ca\,\textsc{ii} IR triplet form in the fast outer layers, whereas weaker lines like Fe\,\textsc{ii} and Mg\,\textsc{ii} arise deeper in the ejecta where the velocity is lower \citep{Dessart2015}.
After approximately $20\,\mathrm{days}$, the He\,\textsc{i}\,$\lambda5876$ velocity exhibits a gradual increase, which may be attributed to blending with Na\,\textsc{i}\,D or the presence of a complex, multi-component velocity structure rather than a single kinematic component. Since we modeled each absorption trough with a single Gaussian component, the inferred velocities may be biased in cases of line blending, particularly for H$\alpha$ and He\,\textsc{i}. Typical uncertainties on the velocity measurements are $\sim 1{,}000\ \mathrm{km\,s^{-1}}$.
For lines that may suffer from blending, we did not attempt deblending and instead fit a single Gaussian profile (e.g., H$\alpha$ potentially blended with Si lines, and He\,\textsc{i}\,$\lambda5876$ with Na\,\textsc{i}\,D).

\subsection{Nebula phase spectrum}

After approximately $70\,\mathrm{days}$, the SN becomes optically thin and enters the nebular phase, exhibiting prominent [O\,\textsc{i}] \(\lambda\lambda6300,6363\), [Ca\,\textsc{ii}] \(\lambda\lambda7291,7324\), and the Ca\,\textsc{ii} infrared triplet \(\lambda\lambda8498,8542,8662\) emission lines. Figure~\ref{fig:nebula} highlights the profile evolution of these lines. All three lines display a pronounced asymmetric profile with double peaks. Such structures are commonly interpreted as signatures of large-scale ejecta asphericity and/or differential internal extinction by newly formed dust within the ejecta \citep{Maeda2008,Taubenberger2009,Bevan2016,Pandey2021,Kumar2025}. These diagnostics provide leverage on the explosion geometry and the onset of dust formation. We further note that the emission of the Ca\,\textsc{ii} near-infrared triplet is strongly concentrated in the $\lambda8662$ component. This behaviour is naturally understood if the triplet remains optically thick at these epochs: resonance scattering and line interlocking among the three fine-structure components redistribute photons absorbed in the $\lambda8498$ and $\lambda8542$ lines into the $\lambda8662$ transition, as shown by detailed radiative-transfer calculations for SN ejecta \citep[e.g.][]{Jerkstrand2017}.

\begin{figure}
    \centering
    \includegraphics[width=1\linewidth]{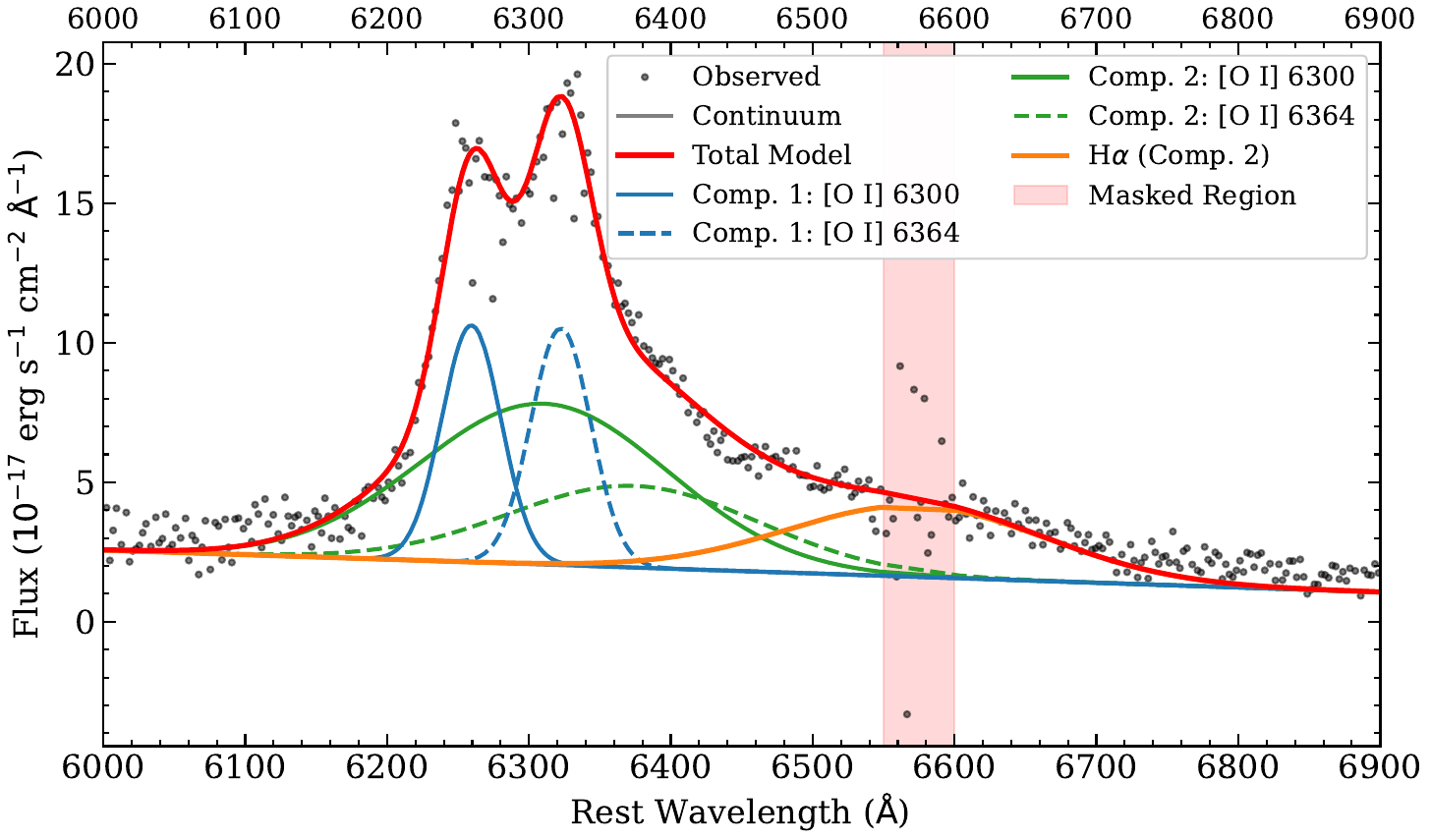}
    \includegraphics[width=1\linewidth]{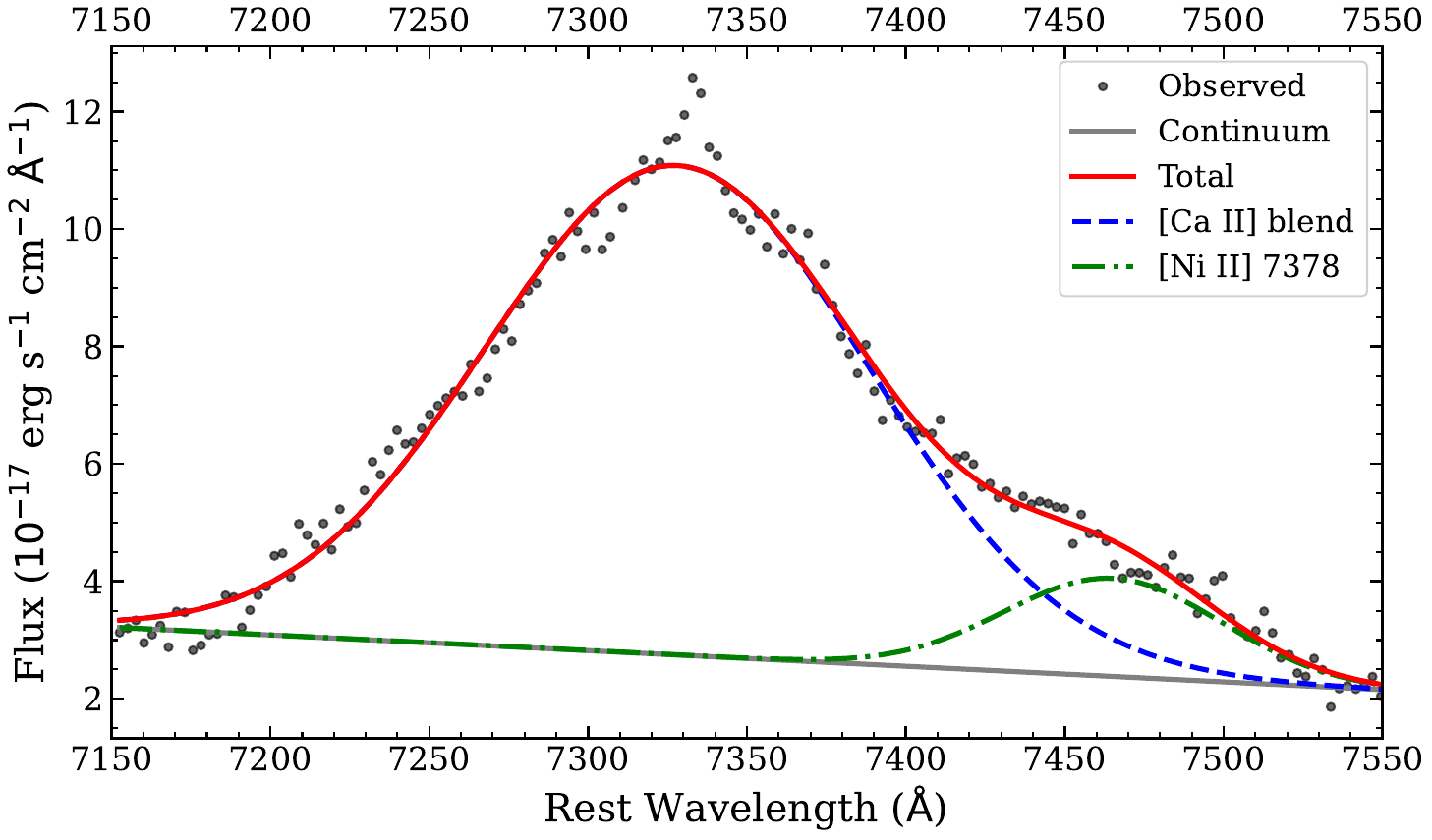}
    \caption{Multi-component fits to nebular emission features of SN~2024aecx. \textit{Top:} the [O\,\textsc{i}] $\lambda\lambda6300,6363$ doublet modeled with two kinematic components that share $v_{\rm shift}$ and FWHM; the wavelength range affected by narrow H$\alpha$ is masked (shaded band). \textit{Bottom:} simultaneous fit to the [Ca\,\textsc{ii}] $\lambda\lambda7291,7324$ blend and [Ni\,\textsc{ii}] $\lambda7378$. Black points show the data, the thin gray curve the pseudo-continuum, the red curve the total model, and the colored dashed/dotted curves the individual components as indicated in the legends. Best-fit parameters are listed in Table~\ref{tab:emission_lines}.}

    \label{fit_emission}
\end{figure}

To probe the geometry and structure of the ejecta of SN~2024aecx, we performed a multi-component fit to the [O\,\textsc{i}] \(\lambda\lambda6300, 6363\) and [Ca\,\textsc{ii}] \(\lambda\lambda7291, 7324\) doublets in the nebular-phase spectrum obtained with the GTC at $t = +123.5$\,days. The results of this fit are presented in Fig.~\ref{fit_emission} and Tab.~\ref{tab:emission_lines}.

Prior to fitting, the spectrum was corrected for the host galaxy's redshift ($z=0.002665$) and dereddened using a foreground extinction of $A_V=1.34$\,mag with $R_V=3.1$, following the extinction law of \cite{CCM89}. For the [O\,\textsc{i}] doublet, a local linear continuum, established by connecting the endpoints of the fitting region, was first subtracted. The resulting emission profile was subsequently modeled with a two-component Gaussian function. For each component, the profiles of the \(\lambda6300\) and \(\lambda6363\) lines were kinematically tied, sharing the same velocity shift and Full Width at Half Maximum (FWHM), while their wavelength separation was fixed at the laboratory value. Allowing for the possibility that the doublet is not entirely optically thin, the flux ratio \(I(\lambda6300)/I(\lambda6363)\) for each component was treated as a free parameter. The best fit reveals a narrow, blueshifted component (comp.~1 in Fig.~\ref{tab:emission_lines}; velocity $v = -1945.23 \pm 24.85$\,${\rm km\,s^{-1}}$, FWHM $= 2354.82 \pm 55.77$\,${\rm km\,s^{-1}}$) and a broader, slightly redshifted component (comp.~2 in Fig.~\ref{tab:emission_lines}; $v = 475.84 \pm 234.36$\,${\rm km\,s^{-1}}$, FWHM $= 9770.33 \pm 268.43$\,${\rm km\,s^{-1}}$). A single Gaussian was also included in the model to account for faint, underlying H\,$\alpha$ and neighboring [N\,\textsc{ii}] emissions.

A similar procedure was applied to the [Ca\,\textsc{ii}] doublet. After subtracting a local linear background, we found that the two lines of the doublet were significantly broadened and blended. Consequently, they were modeled with a single Gaussian profile. An additional Gaussian was incorporated to fit the adjacent [Ni\,\textsc{ii}] \(\lambda7378\) emission line. The best-fit model for the [Ca\,\textsc{ii}] feature yields a velocity of $v = 829.99 \pm 31.20$\,${\rm km\,s^{-1}}$ and an FWHM of $5845.00 \pm 106.17$\,${\rm km\,s^{-1}}$.

Therefore, our analysis reveals a complex kinematic structure in the nebular spectra of SN~2024aecx. The [O\,{\sc i}] $\lambda\lambda$6300, 6364 doublet is best described by a two-component model, whereas the [Ca\,{\sc ii}] $\lambda\lambda$7291, 7324 doublet is well-represented by a single component. We associate the single [Ca\,{\sc ii}] profile with the dominant, low-velocity, broad, and redshifted component of the [O\,{\sc i}] profile (hereafter Component 2). Together, these features trace the bulk of the SN ejecta, suggesting a systematic redshift of the inner explosion region. Crucially, the FWHM of the [O\,{\sc i}] Component 2 is substantially broader than that of the [Ca\,{\sc ii}] profile. Assuming homologous expansion ($v \propto r$), this velocity difference implies a chemically stratified ejecta, where the region of calcium synthesis is confined to the lower-velocity, inner core, while the oxygen-rich material extends to higher velocities in the outer layers. This inferred "inside-out" nucleosynthetic structure is fully consistent with the canonical predictions of CCSN explosion models~\citep{Woosley2007}.

In addition, we identify a distinct, high-velocity, blueshifted component (hereafter Component 1) exclusively in the [O\,{\sc i}] lines. We interpret this feature as a discrete clump or knot of nearly pure oxygen-rich material ejected with a significant line-of-sight velocity component towards the observer. Its large blueshift indicates its high velocity, while its comparatively narrow FWHM suggests a small internal velocity dispersion, consistent with the kinematic signature of a coherent, localized structure. Such high-velocity features, often interpreted as compelling evidence for large-scale ejecta asymmetries, have been observed in the nebular spectra of other CCSNe, a notable example being SN\,2023ixf~\citep{Kumar2025,Michel2025}.

The flux ratio of the [O\,{\sc i}] $\lambda\lambda$6300, 6363 doublet to the [Ca\,{\sc ii}] $\lambda\lambda$7291, 7324 doublet is regarded as a potential diagnostic of the progenitor mass for SESNe.
The underlying principle is that the bulk of the oxygen is synthesized during pre-SN hydrostatic burning phases, the yield of which is highly dependent on the progenitor's initial mass.
In contrast, calcium is predominantly produced during explosive silicon burning, and its synthesized mass is predicted to be relatively insensitive to the progenitor mass over a wide range.
Consequently, a higher [O\,{\sc i}]/[Ca\,{\sc ii}] flux ratio is expected to correspond to a more massive progenitor \citep{Jerkstrand2017}.
For this SN, we measure a ratio of $2.21 \pm 0.15$.
A comparison with the grid of nebular-phase spectral models from \citet{Jerkstrand2015} places our measurement between their 13\,$M_{\odot}$ and 17\,$M_{\odot}$ progenitor models (referred to as m13 and m17, respectively).
This result is consistent with a progenitor of moderate mass, which for a SESN likely implies origin in a binary system.
However, it is important to note that the utility of the [O\,{\sc i}]/[Ca\,{\sc ii}] ratio as a robust mass indicator has been contested. Some studies have found, for instance, a weak or non-existent correlation between this ratio and the total ejecta mass derived from light-curve modeling, casting doubt on its direct correspondence with the progenitor's zero-age main-sequence mass \citep[e.g.,][]{Prentice2022}.

\begin{table*}
\centering
    \caption{Emission-line fit parameters for the nebular phase spectrum of SN~2024aecx. For each transition we list the velocity shift $v_{\rm shift}$, the FWHM, and the integrated flux. The [O\,\textsc{i}] $\lambda\lambda6300,6363$ doublet is fitted simultaneously with two kinematic components that share $v_{\rm shift}$ and FWHM; [Ca\,\textsc{ii}] refers to the $\lambda\lambda7291,7324$ blend; [Ni\,\textsc{ii}] $\lambda7378$ is modeled as a single component. The corresponding spectral fit is shown in Fig.~\ref{fit_emission}.}
\label{tab:emission_lines}
\begin{tabular}{lccc}
\hline\hline
Line & $v_{\rm shift}$ (km\,s$^{-1}$) & FWHM (km\,s$^{-1}$) & Flux ($10^{-17}$)$^a$ \\
\hline
{[O\,\textsc{i}] \(6300\) Comp.\,1} & $-1945.23 \pm 24.85$ & $2354.82 \pm 55.77$ & $444.49 \pm 20.94$ \\
{[O\,\textsc{i}] \(6363\) Comp.\,1} & $-1945.23 \pm 24.85$ & $2354.82 \pm 55.77$ & $448.97 \pm 23.46$ \\
{[O\,\textsc{i}] \(6300\) Comp.\,2} & $475.84 \pm 234.36$  & $9770.33 \pm 268.43$ & $1262.87 \pm 133.31$ \\
{[O\,\textsc{i}] \(6363\) Comp.\,2} & $475.84 \pm 234.36$  & $9770.33 \pm 268.43$ & $647.37 \pm 114.30$ \\
{[Ca\,\textsc{ii}] blend}           & $829.99 \pm 31.20$   & $5845.00 \pm 106.17$ & $1267.70 \pm 30.26$ \\
{[Ni\,\textsc{ii}] \(7378\)}        & $3548.21 \pm 133.57$ & $3274.20 \pm 11.76$  & $144.93 \pm 11.06$ \\
\hline
\end{tabular}
\tablecomments{$^a$ Flux units in erg\,s$^{-1}$\,cm$^{-2}$.}
\end{table*}

\section{Modeling of the Progenitor and Explosion Parameters}
\label{sec:model}

\begin{deluxetable*}{l l c c c}

\tablecaption{Key Model Parameters and Priors for SN~2024aecx\label{tab:sn2024aecx_params}}
\tablewidth{0pt}

\tablehead{
  \colhead{Parameter} &
  \colhead{Definition} &
  \colhead{Units} &
  \colhead{Default Prior} &
  \colhead{SN~2024aecx}
}
\startdata
$M_{\mathrm{e}}$         & Extended H-rich envelope mass     & $M_{\odot}$              & L[$10^{-3}$, $0.5$]          & $0.04\pm0.01$ \\
$R_{\mathrm{e}}$         & Extended envelope radius          & $R_{\odot}$              & L[$10$, $10^{4}$]            &  $66.75^{+25.47}_{-25.93}$ \\
$E_{\mathrm{e}}$         & Energy of  extended envelope      & $10^{49}\,\mathrm{erg}$  & L[$0.01$, $20$]              & $0.96^{+0.97}_{-0.36}$ \\
$M_{\mathrm{ej}}$        & Ejecta mass                       & $M_{\odot}$              & L[$0.3$, $10$]               & $1.55^{+0.18}_{-0.14}$ \\
$M_{\mathrm{Ni}}$        & Synthesized $^{56}$Ni mass        & $M_{\odot}$              & L[$0.01$, $0.3$]             & $0.09\,\pm\,0.01$ \\
$t_{0}$                  & Explosion epoch$^{a}$             & day                      & U[$0$, $5$]                  & $0.02^{+0.02}_{-0.01}$ \\
\enddata
\tablecomments{L = log-uniform prior; U = uniform prior. The \textit{Default Prior} column lists the priors adopted in our MCMC analysis of the bolometric light curve. $^{a}$Defined as the time offset between the explosion epoch and the discovery date.}
\end{deluxetable*}

The evolution of a SN light curve in the hours and days immediately following the explosion is a powerful diagnostic of the progenitor star's structure and recent mass-loss history. SN~2024aecx is a compelling example, discovered shortly after its explosion, with an optical light curve that exhibits a prominent double-peaked morphology (Section~\ref{sec:lightcurve}; Fig.~\ref{fig:bol}). This structure is characteristic of an initial, rapidly fading peak from shock-cooling emission from an extended stellar envelope, followed by a broader, slower peak powered by the radioactive decay of $^{56}\mathrm{Ni}$ and $^{56}\mathrm{Co}$. Such double-peaked light curves, famously observed in SNe like SN~1993J and SN~2011dh, are often interpreted as evidence for progenitors with extended, low-mass hydrogen envelopes \citep[e.g.,][]{Arcavi2011, Bersten2012}. And statistical study based on ATLAS further indicates that nearly half of Type~IIb SNe exhibit this kind of early double-peaked behavior, implying that such progenitor configurations may be relatively common within this class \citep{Ayala2025}. 

To interpret these features in SN~2024aecx, we model its bolometric light curve using a semi-analytic framework that incorporates two distinct physical components. First, the early peak is attributed to shock-cooling emission, which we calculate using the analytic prescriptions of \citet{Piro2021} to connect the peak's luminosity and duration to the envelope's mass and radius. Second, the main peak is powered by radioactive heating, which is modeled with the classical diffusion formalism of \citet{Arnett1982} to treat the thermalization of energy from the $^{56}\mathrm{Ni} \rightarrow ^{56}\mathrm{Co} \rightarrow ^{56}\mathrm{Fe}$ decay chain. This two-component framework provides a computationally efficient yet physically grounded description of the light curve, enabling a direct connection between its observed evolution and the underlying properties of the progenitor and explosion.

\begin{figure}
    \centering
    \includegraphics[width=1\linewidth]{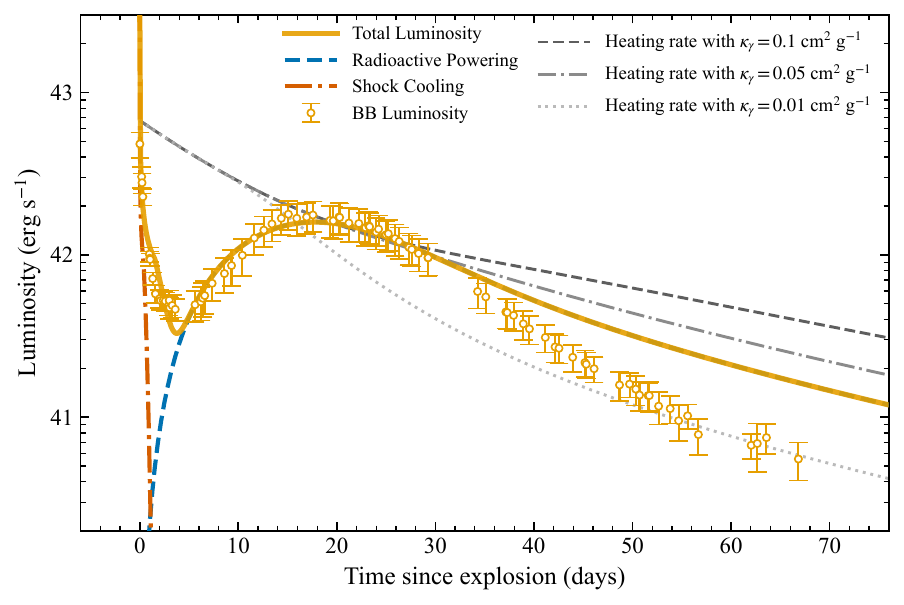}
\caption{Bolometric light curve of SN~2024aecx (orange circles) compared with the best-fit two-component model. The solid orange curve shows the total luminosity, decomposed into radioactive powering emission (blue dashed line) and shock-cooling emission (brown dash–dotted line). The gray curves show the instantaneous radioactive heating rate for different assumed $\gamma$-ray opacities,
$\kappa_\gamma = 0.1$ (dark gray dashed), $0.05$ (medium gray dash–dotted),
and $0.01~\mathrm{cm^{2}\,g^{-1}}$ (light gray dotted). The model reproduces both the early cooling peak and the main radioactive peak, while the late-time decline is steeper than the corresponding heating rate.}
    \label{fig:lc_fitting}
\end{figure}

\begin{figure*}
    \centering
    \includegraphics[width=1\linewidth]{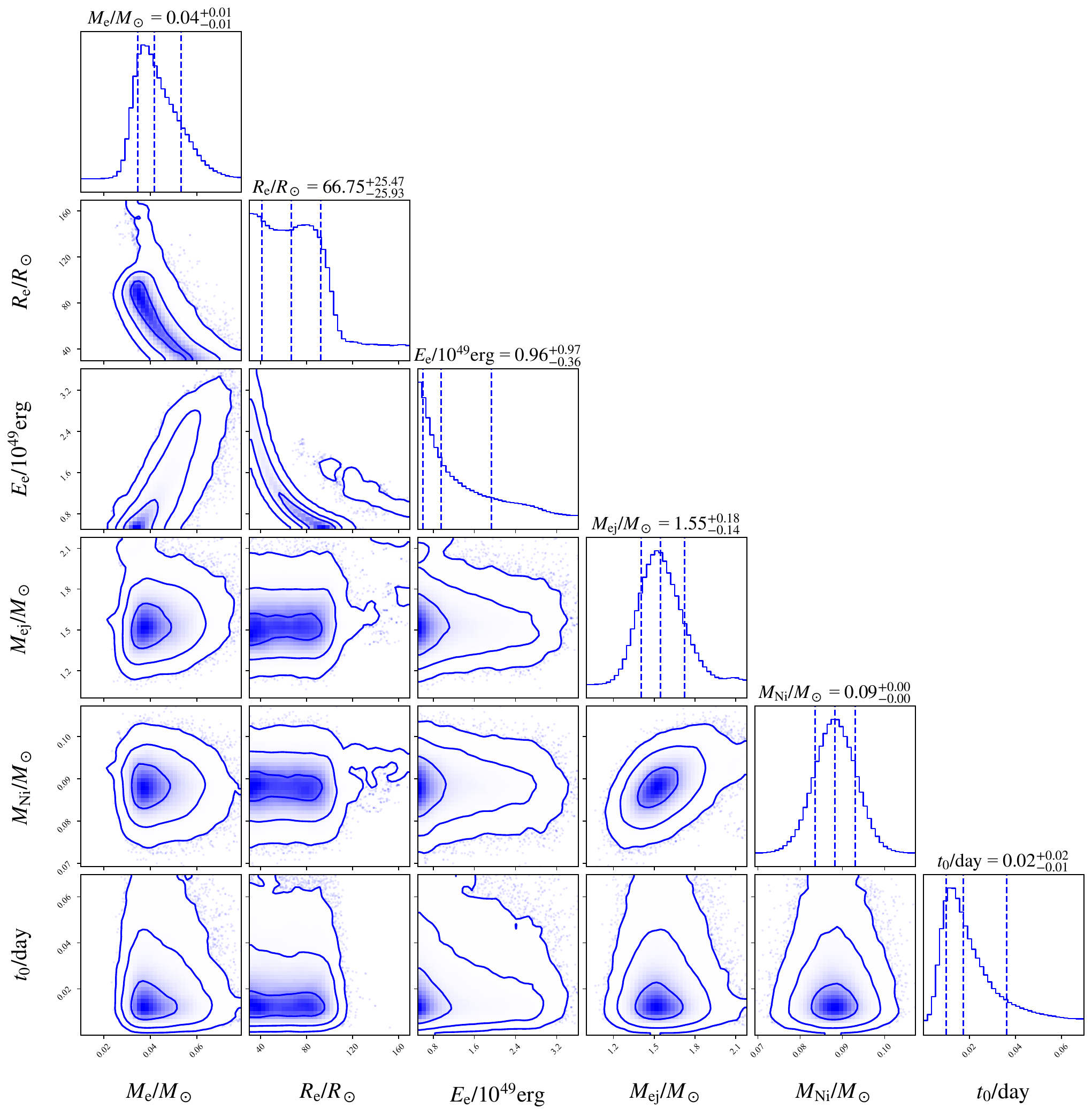}
\caption{The posterior probability distributions and covariances for the key physical
parameters of SN 2024aecx from our MCMC analysis.
  }
    \label{fig:fitting_corner}
\end{figure*}

To quantitatively constrain the physical parameters governing the light curve of SN~2024aecx, we performed a comprehensive Bayesian analysis using a Markov Chain Monte Carlo (MCMC) approach. The model parameters, their corresponding prior distributions, and the resulting best-fit values are summarized in Table~\ref{tab:sn2024aecx_params}. The fit to the bolometric light curve is presented in Figure~\ref{fig:lc_fitting}. The resulting posterior probability distributions for the key physical parameters of our model are presented in the corner plot in Figure~\ref{fig:fitting_corner}. As an external spectroscopic constraint, Fe\,II absorption features around maximum light indicate a photospheric velocity of \(\sim6500\,\mathrm{km\,s^{-1}}\) (Section~\ref{sec:spectroscopy}; Fig.~\ref{fig:velocity_evolution}); accordingly, we adopt \(v_{\rm ej}=6500\,\mathrm{km\,s^{-1}}\) in the fit.

Our analysis yields an extended envelope mass of \( M_{\mathrm{e}} = 0.04\pm0.01\, M_{\odot} \), radius \( R_{\mathrm{e}} = 66.75^{+25.47}_{-25.93}\, R_{\odot} \), and characteristic energy \( E_{\mathrm{e}} = 0.96^{+0.97}_{-0.36} \times 10^{49} \, \mathrm{erg} \) for the shock cooling component. For the radioactive component, we infer an ejecta mass of \( M_{\mathrm{ej}} = 1.55^{+0.18}_{-0.14} \, M_{\odot} \), and a synthesized nickel mass of \( M_{\mathrm{Ni}} =0.09\,\pm\,0.01\, M_{\odot} \). These values are consistent with a moderately extended, low-mass hydrogen envelope and a highly energetic core-collapse explosion. These properties closely match expectations for a binary-stripped progenitor, where Roche-lobe overflow removes most of the hydrogen envelope. Indeed, all directly detected IIb progenitors to date are linked to binaries, and our results provide further support for this scenario \citep{Maund2004,Niu2024,Niu2025}.

As shown in Figure~\ref{fig:lc_fitting}, the bolometric light curve of SN~2024aecx is well reproduced by our two-component model: at very early times ($t\lesssim5$~d) the luminosity clearly exceeds the radioactive heating curves, confirming that the first peak is dominated by shock cooling of the extended envelope rather than by $^{56}$Ni decay. Around the main maximum ($\sim10$–30~d), the total luminosity is matched when adopting a $\gamma$-ray opacity of $\kappa_\gamma \simeq 0.05~\mathrm{cm^2\,g^{-1}}$, indicating efficient $\gamma$-ray trapping in the bulk of the ejecta. At later epochs ($t \gtrsim 40$~d), however, the observed luminosity declines more rapidly than predicted by the radioactive heating model, even after including the effect of $\gamma$-ray leakage from the ejecta \citep{Clocchiatti1997,Maeda2003}. This suggests that the emission is no longer fully controlled by the instantaneous $^{56}$Co decay power and that additional processes—such as early dust formation \citep{Gall2014}, incomplete positron trapping \citep{Milne2001}, or asymmetric ejecta that enhance high-energy photon escape \citep{Maeda2008}—may contribute to the steep decline. Such behavior is consistent with the relatively low ejecta mass inferred from our modeling and with the rapid post-peak evolution commonly seen in SNe~IIb .
Even in comparison with such SESNe (e.g., SN~2008ax \citep{Taubenberger2011} and SN~2011ei \citep{Milisavljevic2013}; see also \citep{Maeda2008}), SN~2024aecx still declines unusually fast (Fig.~\ref{fig:comparison}); a similar trend has been reported by \citet{Zou2025}, suggesting that $\gamma$-ray leakage alone may not fully explain this phenomenon.

Given the derived parameters, SN~2024aecx likely originated from a progenitor with a moderately extended, low-mass hydrogen envelope and an energetic core-collapse explosion. Future multi-wavelength follow-up, especially in the infrared and polarimetric domains, will be essential to test the dust-formation and ejecta-asymmetry scenarios.

In comparison with the results of \citet{Zou2025}, our analysis of SN\,2024aecx yields broadly consistent extinction estimates and a similar velocity evolution of spectral features. However, our dataset benefits from a denser spectroscopic cadence, the inclusion of nebular-phase spectra, and more extensive multi-band photometry. Our modeling recovers the same $^{56}$Ni mass \( M_{\mathrm{Ni}} \simeq 0.09 \, M_{\odot} \), while constraining the envelope mass to \( M_{\mathrm{e}} = 0.04\pm{0.01}\, M_{\odot} \), which falls within the range of $\sim$0.03–0.24\,\( M_{\odot} \) reported by \citet{Zou2025}, but with significantly tighter uncertainties and a smaller inferred envelope radius. We also derive a larger ejecta mass of \( M_{\mathrm{ej}} = 1.55^{+0.18}_{-0.14}\, M_{\odot} \), compared to the $\sim$0.7\,\( M_{\odot} \) obtained by \citet{Zou2025}. Further detailed investigations of the photometric and spectroscopic evolution of this SN are warranted.

\section{Summary}
\label{sec:conclusion}

We presented a comprehensive UV-optical study of the Type~IIb SN\,2024aecx in NGC\,3521. SN\,2024aecx was discovered within $\sim$1\,d from the last non-detections, and our follow-up campaign started from as early as $t=0.26$\,d after discovery and continued out to the nebular phase.

With the TRGB method we obtain a new distance of $D=11.3\pm1.1$\,Mpc to the host galaxy. VLT/MUSE spectroscopy of the co-spatial H\,\textsc{ii} region indicates a near-solar abundance of $12+\log(\mathrm{O/H})=8.49\pm0.18$ (O3N2). We also derive a host extinction of $A_V^{\rm Host}\simeq1.2$\,mag toward SN\,2024aecx.

The light curves display a short-lived early peak, attributable to the cooling of a shock-heated extended surface layer, followed by a secondary maximum and an unusually rapid post-peak decline. While the peak luminosity fall within the typical range for this SN class, both the shock-cooling phase and the subsequent decline after the secondary maximum proceed at an exceptionally fast pace.

The earliest spectra at $t=0.3-1.2$\,d display a blue, nearly featureless continuum. From $t=2.3$\,d, P-Cygni features emerge and we derived the velocity evolution of H\,$\alpha$, He\,\textsc{i}\,$\lambda5876$, Mg\,\textsc{ii}\,$\lambda4481$, the Ca\,\textsc{ii} infrared triplet, and the Fe\,\textsc{ii} triplet. We took three nebular-phase spectra at $t=71.5, 123.5, 158.5$\,d. The [O \textsc{i}] and [Ca \textsc{ii}] lines exhibit asymmetric, double-peaked profiles, which may arise from asphericity and/or early dust formation.

Our modeling of the nebular-phase spectrum of SN\,2024aecx reveals a highly asymmetric explosion geometry. The observed line profiles can be well-reproduced by a model consisting of a prominent, blueshifted oxygen-rich clump of ejecta, superimposed on a slightly redshifted bulk component. We measure the flux ratio of the [O\,{\sc i}]~$\lambda\lambda$6300, 6364 doublet to the [Ca\,{\sc ii}]~$\lambda\lambda$7291, 7324 doublet to be approximately 2. This value is consistent with those observed in other CCSNe, suggesting that SN\,2024aecx likely arose from a progenitor of moderate mass, possibly in a binary system.

A two-component model combining shock cooling and $^{56}$Ni heating reproduces the overall double-peaked morphology of the bolometric light curve. Inference favors a dilute, extended H-rich envelope (\( M_{\mathrm{e}} = 0.04\pm0.01\, M_{\odot} \), \( R_{\mathrm{e}} = 66.75^{+25.47}_{-25.93} \, R_{\odot} \)) and a SE explosion with an ejecta mass of \( M_{\mathrm{ej}} = 1.55^{+0.18}_{-0.14} \, M_{\odot} \) and \( M_{\mathrm{Ni}} = 0.09\,\pm\,0.01\, M_{\odot} \). The late-time decline proceeds faster than expected, suggesting enhanced $\gamma$-ray leakage and/or dust formation. The nebular-phase spectra and light-curve modelling both suggest that it most likely originated from an intermediate-mass binary progenitor system.

SN\,2024aecx is one of the well-sampled Type~IIb SNe: at a very close distance, discovered at an exceptionally early phase, and exhibiting one of the most prominent shock-cooling flashes among Type~IIb SNe. The combination of high-cadence photometric and spectroscopic follow-up observations along with the well-characterized environment provides stringent constraints on the progenitor and its explosion. The data and analysis presented here offer a high-quality reference for shock-cooling physics and progenitor-envelope coupling in Type~IIb SNe. Future near-IR photometry and mid-IR spectroscopy can directly test dust formation; radio/X-ray observations will probe circumstellar interaction and recent mass loss; and nebular radiative-transfer modeling and spectropolarimetry will further quantify ejecta geometry.

\begin{acknowledgments}

We are grateful to Prof. Luc Dessart  and Prof. Anders Jerkstrand for their valuable discussions and suggestions regarding the nebular-phase spectra. We also thank Prof. Nancy Elias-Rosa for her helpful discussions during the course of this work. We thank the anonymous referee for the time devoted to reviewing our manuscript and for the valuable comments.

NCS is funded by the Strategic Priority Research Program of the Chinese Academy of Sciences Grant No. XDB0550300, the National Natural Science Foundation of China Grants No.12303051 and No. 12261141690, and the China Manned Space Program No. CMS-CSST-2025-A14. FP acknowledges support from the Spanish Ministerio de Ciencia, Innovaci\'on y Universidades (MICINN) under grant numbers PID2022-141915NB-C21. KM, BA and DK acknowledge the support from the BRICS grant DST/ICD/BRICS/Call-5/CoNMuTraMO/2023 (G) funded by the Department of Science and Technology (DST), India. LL and YZ acknowledge the support from the National Natural Science Foundation of China (grant Nos 12303047) and Natural Science Foundation of Hubei Province (2023AFB321). ZG acknowledge the support supported from the China-Chile Joint Research Fund (CCJRF No.2301) and the Chinese Academy of Sciences South America Center for Astronomy (CASSACA) Key Research Project E52H540301. ZG, is funded by ANID, Millennium Science Initiative, AIM23-001. WXL acknowledge the supports from National Key R\&D Program of China (grant Nos. 2023YFA1607804, 2022YFA1602902, 2023YFA1607800, 2023YFA1608100), the National Natural Science Foundation of China (grant Nos. 12120101003, 12373010, 12173051, and 12233008), China Manned Space Project (No. CMS-CSST-2025-A06) and the Strategic Priority Research Program of the Chinese Academy of Sciences with Grant Nos. XDB0550100 and XDB0550000. LZW is sponsored by the National Natural Science Foundation of China (NSFC) grant No. 12573050,  the Chinese Academy of Sciences South America Center for Astronomy (CASSACA) Key Research Project E52H540301, and in part by the Chinese Academy of Sciences (CAS) through a grant to the CASSACA. JDL acknowledges support from a UK Research and Innovation Future Leaders Fellowship (MR/T020784/1). BCW acknowledges support from the National Key R\&D Program of China (grant Nos. 2023YFA1609700, 2023YFA1608304), the National Natural Science Foundation of China (NSFC; grant Nos. 12090040, 12090041, 12403022), and the Strategic Priority Research Program of the Chinese Academy of Sciences (grant Nos. XDB0550000, XDB0550100, XDB0550102).

Based on observations made with the Thai Robotic Telescopes under program IDs TRTC12A\_003, TRTC12A\_002, TRTToO\_2024006, and TRTC11C\_007, operated by the National Astronomical Research Institute of Thailand (Public Organization).

We thank the staff at all participating observatories, including the Xinglong 60\,cm and 2.16\,m telescopes, the UCAS-70 telescope, the Devasthal Fast Optical Telescope, the Liverpool Telescope, the Thai Robotic Telescope network, the Las Cumbres Observatory network, the Himalayan Chandra Telescope, the Devasthal Optical Telescope, the Nordic Optical Telescope, and the Gran Telescopio Canarias, for their support during our observing campaigns.

We acknowledge the use of data and services provided by ATLAS, ZTF, GOTO, Swift/UVOT, the Transient Name Server (TNS), the Mikulski Archive for Space Telescopes (MAST), and the ESO Science Archive Facility. This research has made use of NASA’s Astrophysics Data System (ADS). We also acknowledge the use of open-source software packages including \textsc{astropy}, \textsc{AutoPhot}, \textsc{hotpants}, \textsc{IRAF}, and \textsc{PypeIt}.

\end{acknowledgments}

\facilities{DFOT, DOT, GTC, HCT, HST/ACS, LT, NOT, Swift/UVOT, TRT, UCAS-70, VLT/MUSE, XL-60, XL-216}

\software{astropy \citep{2013A&A...558A..33A,2018AJ....156..123A,2022ApJ...935..167A},  
AutoPhot \citep{Brennan2024}, hotpants \citep{Becker2015}, IRAF \citep{Tody1986},
PypeIt \citep{Prochaska2020}}



\bibliographystyle{aasjournal}
\bibliography{sample701}{}



\end{document}